\documentclass[10pt,tightenlines,eqsecnum,floats,aps,amsmath,
amssymb,nofootinbib,prd,showpacs]{revtex4}
\usepackage{amsmath,amsthm,latexsym,amssymb,amsfonts}
\usepackage[T1]{fontenc}
\usepackage{enumerate}
\usepackage{graphicx}
\usepackage{calc}
\usepackage{relsize}

\newcommand{\R}{{\mathbb R}}

\newcommand{\gr}{{\rm gr}}

\newcommand{\Hil}{\mathcal{H}}

\newcommand{\re}{\mathbb{R}}

\input cyracc.def
\newfam\cyrfam
\font\tencyr=wncyr10
\font\sevencyr=wncyr7
\font\fivecyr=wncyr5
\def\cyr{\fam\cyrfam\tencyr\cyracc}
\textfont\cyrfam=\tencyr \scriptfont\cyrfam=\sevencyr
\scriptscriptfont\cyrfam=\fivecyr
\newcommand{\ppi}{{\cyr p}}

\newcommand{\id}{\mathbb{I}}

\newcommand{\rd}{{\rm d}}
\newtheorem{lm}{Lemma}
\newtheorem{df}[lm]{Definition}

\begin{document}

\title{Quantum constraints, Dirac observables and evolution:\\ group averaging
versus Schr\"odinger picture in LQC}

\author{Wojciech Kami\'nski${}^{1}$}
\email{wkaminsk@fuw.edu.pl}
\author{Jerzy Lewandowski${}^{1}$}
\email{lewand@fuw.edu.pl}
\author{Tomasz Paw{\l}owski${}^{2,1}$}
\email{tomasz@iem.cfmac.csic.es}

\affiliation{
  ${}^{1}$Instytut Fizyki Teoretycznej, Uniwersytet Warszawski,
  ul. Ho\.{z}a 69, 00-681 Warszawa, Poland\\
  ${}^{2}$Instituto de Estructura de la Materia, CSIC, 
  Serrano 121, 28006 Madrid, Spain}

\begin{abstract} \noindent{\bf Abstract\ }
A general quantum constraint of the form 
${\hat C}\ =\ -\frac{\partial^2}{\partial T^2} \otimes \hat{B}-\hat{\id}\otimes \hat{H}$ 
(realized in particular in Loop Quantum Cosmology models)
is studied. Group Averaging is applied to define the Hilbert space of solutions and the relational 
Dirac observables. Two cases are considered. In the first case, the spectrum of the operator 
$\frac{1}{2}\ppi^2\hat{B} -\hat{H}$ is assumed to be discrete. The quantum theory defined by 
the constraint takes the form of a Schr\"odinger-like quantum mechanics with a generalized 
Hamiltonian  $\sqrt{\hat{B}^{-1}\hat{H}}$. In the second case, the spectrum is absolutely 
continuous and some peculiar asymptotic properties of the eigenfunctions are assumed. 
The resulting Hilbert space and the dynamics are characterized by a continuous family of 
the Schr\"odinger-like quantum theories. However, the relational observables mix different 
members of the family. Our assumptions are motivated by new Loop Quantum Cosmology models 
of quantum FRW spacetime. The two cases considered in the paper correspond  to the negative 
and, respectively, positive  cosmological constant. Our results should be also applicable in 
many other general relativistic contexts.
\end{abstract}


\maketitle

\def\be{\begin{equation}}
\def\ee{\end{equation}}
\def\ba{\begin{eqnarray}}
\def\ea{\end{eqnarray}}
\def\lp{{\ell}_{\rm Pl}}
\def\g{\gamma}

\section{Motivation to understand quantum constraints}\label{sec:intro}

To construct a canonical quantum theory out of the classical theory
with constraints, like quantum gravity, one usually employs Dirac program, in
which the physical Hilbert space is built out of the space of solutions to the
constraints represented as quantum operators acting in the kinematical space.
The formulation of the program however allows for a certain amount of ambiguity
in performing particular steps. One of its sources is the identification of the
precise constraint condition, that is the exact mechanism, via which the
constraint operator selects the physical Hilbert space. Another one is
the formulation (and meaning) of the physical evolution of the system. A
proposal which in many examples provides a systematic way to address the first
issue is known as \emph{Group Averaging} \cite{Marolf2,Marolf}. That framework 
combined with the idea of ``partial'' or ``relational'' observables 
\cite{Rovelli1,rel-obs}, provides also in a precise way the solution to the second 
problem. Therefore it seems to be the most promising tool to complete the task of 
constructing canonical quantum gravity.

One of the formulations of such theory being particularly close to the point of
completion is Loop Quantum Gravity \cite{Rovelli,A-LQG,Thiemann} coupled with Brown-Kuchar dust fields 
\cite{lqg-bk}. There one selects one of the dust fields as internal time and deparametrizes the
theory with respect to it. The Hamiltonian constraint is reformulated as the
Schr\"odinger equations generating an evolution with respect to selected time.
On the other hand, one can apply the methods of group averaging directly to the
constraints in their original form. That possibility in turn opens the room for a question, whether
both physical pictures resulting from these approaches do necessarily coincide.

The suggestion, that the answer to this question might be nontrivial
comes from Loop Quantum Cosmology \cite{LQC,A-LQC,LQC-ABL} which constitutes a good testing ground for
Loop Quantum Gravity. The LQC models share more common features with LQG than
any other example \cite{A-LQC,klp-aspects}. At the same time they are technically simple enough to study
the mathematical properties of quantum constraints \cite{kp-constr}, the structure of physical
Hilbert space, the quantum solutions and observables \cite{APS,APS-imp}. In the context of
considered problem the signal of possible inequivalence  (at least in some
situations) shows up at the level of the basic properties of the operators
involved in each approach. Indeed, the studies of the models of
Friedman-Robertson-Walker universes with positive cosmological constant reveal
\cite{klp-aspects,ap-posL,kp-posL} that, while the quantum Hamiltonian
constraint (the substrate for group averaging) operator is essentially
self-adjoint, the evolution operator in the Schr\"odinger picture is not. In
consequence the two approaches seem to give different answers even to the
question whether the defined physical evolution is unique.

We address the issue of equivalence between group averaging and Schr\"odinger
picture in this article. For the universality we focus our attention on an
(abstract) constraint, whose structure and certain properties (relevant for this
problem) coincide with the ones of the Hamiltonian constraint describing FRW
model with nonvanishing cosmological constant in LQC \cite{bp-negL,ap-posL,kp-posL}. 
This makes the results (perhaps after suitable generalization) extendable to more general 
cases, potentially including in particular LQG with Brown-Kuchar dust fields.

To start with, let us define, what we understand in our studies as the
Schr\"odinger picture. For that let us consider a quantum constraint
operator
\be
  \hat{C}_1\ =\ \frac{1}{i}\frac{\partial}{\partial T}\otimes \hat{\id}-\hat{\id}\otimes \hat{H}\label{1}
\ee
defined in the Hilbert space  $L^2(\mathbb{R})\otimes {\cal H}$ where $\frac{\partial}{\partial T}$ is the derivative operator. If we write the action of the quantum constraint operator as
\begin{equation}
  (\hat{C}\psi)(T)\ =\ \frac{1}{i}\frac{\partial}{\partial T}\psi(T)\ -\ \hat{H}\psi(T) \ ,
\end{equation}
then everybody will agree that a reasonable definition of solution to $\hat{C}_1$
is:\\
a function
\begin{equation}
  \mathbb{R}\ni T\ \mapsto\ \psi(T)\in{\cal H} \ ,
\end{equation}
such that
\be 
  \frac{1}{i}\frac{\partial}{\partial T}\psi(T)\ =\ \hat{H}\psi(T) \ . \label{1a}
\ee
The structure of the solutions to ${\hat C}_1$ takes then the structure 
characteristic to the Shr\"odinger quantum mechanics
with the Hamiltonian operator ${\hat H}$ and the Hilbert space
${\cal H}$ \cite{Rovelli1}. An operator defined in ${\cal H}$ (kinematical
observable) defines an operator acting on the solutions of the constraint
(Dirac observable) provided  an instant $T=T_0$ is given.

In the Special Relativity context, a more common example is a
quadratic constraint, that is
\be
  {\hat C}_2\ =\ -\frac{\partial^2}{\partial T^2} \otimes \hat{\id}-\hat{\id}\otimes
  \hat{H} \ .\label{2}
\ee 
which however can be reduced to the previous case
by employing the decomposition onto positive and negative frequency sectors and writing 
\eqref{2} as
\be 
  \frac{1}{i}\frac{\partial}{\partial T}\psi(T)\ 
  =\ \pm \sqrt{\hat{H}}\, \psi(T) \ .\label{2a}
\ee

A further complication  emerges in the General Relativity context, where a quantum 
constraint operator can take the following form,
\be 
  {\hat C}\ =\ -\frac{\partial^2}{\partial T^2} \otimes \hat{B}-\hat{\id}\otimes \hat{H} \ , \label{3}
\ee
where $\hat{B}$ is an  operator in ${\cal H}$. Then, typically one turns the constraint 
into the following equation
\be 
  \frac{1}{i}\frac{\partial}{\partial T}\psi(T)\ =\ \pm \sqrt{\hat{B}^{-1}\hat{H}}\, \psi(T) , \label{3a}
\ee
defined in the Hilbert space obtained from ${\cal H}$ by a suitable change of the scalar 
product \cite{Fulling,APS}. This prescription, to which we will refer to as \emph{the Schr\"odinger 
picture}, has been in particular quite extensively employed in the description of the dynamics of 
LQC models (see \cite{APS-imp,bp-negL,ap-posL,apsv,b1-szulc} and in the modified form adopted to 
polymeric space structure \cite{mmp2}).

On the other hand, there is being developed a more general, systematic treatment 
of quantum constraint expressed by a self-adjoint operator of arbitrary form. The 
space of solutions is defined by the spectral decomposition of the quantum constraint 
operator, the Dirac observables are constructed by using relational observables 
\cite{rel-obs} This is a special, 1-constraint  case of the Group Averaging (GA)
method (or ``rigging map'') \cite{Marolf2,Marolf,Thiemann}. In the simplest cases 
(\ref{1},\ref{2}), the spectral/GA  method is known to give simply (\ref{1a},\ref{2a}). 
Our goal, is  application of the spectral/GA methods to a quantum constraint of the 
form (\ref{3}), derivation of the Hilbert space of the solutions and comparing 
it with the structure provided by the Schr\"odinger picture following from (\ref{3a}). 
We consider two cases, given by two different sets of assumptions. In the first 
case, denoted here as the \emph{discrete} one, the spectrum of the operator 
${\hat H}$ is assumed to be discrete. The results of the group averaging are 
equivalent to those of the Schr\"odinger picture. An advantage in this case 
is that we arrive to the result via systematic application of a quite general 
method. In the second case (denoted as \emph{continuous} one) the spectrum 
of ${\cal H}$ is absolutely continuous. In here we impose some additional 
assumptions concerning the asymptotic behaviour of the eigenfunctions forming the 
basis in the spectral decomposition. In this example, the derived space of 
solutions has a more interesting structure. In particular, it turns out that 
while the right hand side of (\ref{3a}) can not be uniquely defined, the 
result of group averaging is unique. The observables derived systematically via 
the latter method have in this case even more suprising properties. At this point 
however the reader should be aware, that, while the latter case was denoted as continuous, 
the continuity of the spectrum of $\hat{C}$ is not enough to ensure the reported properties.
In particular there exist examples, for which both the spectrum of $\hat{B}^{-1}\hat{H}$ and 
(the part of) the spectrum of $\hat{C}$ are continuous, nonetheless both the operators 
define a unique physical evolution. The asymptotic properties of the basis functions 
play here an essential role.

We introduce the discrete and continuous case by formulating the
assumptions it satisfies, rather then by giving two specific
examples. However, examples do exist and we found them in Loop Quantum
Cosmology, more precisely in the model of the massless scalar field coupled with
the homogeneous, isotropic universe \cite{APS-imp,bp-negL,ap-posL}. The properties 
of the quantum scalar constraint operator depend there on the sign of the
cosmological constant \cite{klp-aspects}. The quantum scalar constraint with the
negative cosmological constant is a specific example of the discrete case,
whereas  the constraint with the positive cosmological constant
provides the original example of the continuous one.  Our attention to 
the problem considered in this article was drawn exactly by certain puzzling 
observation concerning the quantum scalar constraint corresponding to the positive 
cosmological constant. On the one hand, the operator on the right hand side of
$(\ref{3a})$ admits many inequivalent self-adjoint extensions. Each extension defines 
a distinct Hilbert space of solutions and a distinct quantum theory, which however 
provide same physical predictions. On the other hand, the quantum constraint
operator $\hat{C}$ (\ref{3}) has a unique self adjoint extension for
arbitrary cosmological constant \cite{klp-aspects} and via the spectral/GA
method it defines a unique quantum theory. Then the natural question arises:
what is the relation between the solutions according to the
spectral/GA method, and the solutions defined by each self adjoint
extension of the right hand side of (\ref{3a})? Also, how do the Dirac
observables enter those spaces of solutions? The results presented here provide a 
solution to that puzzle.

The paper is organized as follows. We start with a general introduction to the 
Group Averaging in Section \ref{sec:GA-intro} (subsections \ref{GAclass} and 
\ref{GAq}). In the description we smuggle in a somewhat original, generalized 
formulation. Our formula for the relational observables is slightly different 
than that in \cite{rel-obs} and coincides with that of \cite{Marolf2}. 

After the general introduction we discuss in more detail the spectral/GA method in 
a case of a constraint which has the structure considered in our paper, that is the 
constraint characteristic to the model of FRW universe. This provides the starting point 
to the technical part of our paper.

We start it with the summary (in Section \ref{sec:res}) of the results derived in 
the paper. The actual detailed derivations are contained in the following Sections 
\ref{Exp1} and \ref{Exp2} dedicated, respectively, to the discrete and continuous case.
Each of the sections is concluded with an individual short summary and discussion, 
however the main results, their consequences and possible extensions are discussed in 
the concluding Section \ref{sec:concl}.

\section{Group Averaging for a finite dimensional group}\label{sec:GA-intro}

The group averaging procedure introduced in \cite{Marolf} is a powerful and
quite universal method allowing to define physical Hilbert space in constrained
quantum systems as well as provides a way to build Dirac observables. In this
section we present a brief introduction to this procedure, considering it on
two levels: classical and quantum. After general discussion we focus on the systems 
with $1$ constraint, represented by the models of FRW universe studied in LQC.

\subsection{Classical formulation}\label{GAclass} 

Consider a classical theory in a phase space $\Gamma$ equipped with a Poisson bracket 
$\{\cdot,\cdot\}$. Suppose the physical phase space of the system is a submanifold of $\Gamma$ satisfying
\be 
  C_1\ =\ 0 \ ,\quad  \ldots ,\ \ C_d\ =\ 0 \ , \label{clascon}
\ee
where $C_1,...,C_d$ are some real valued functions on $\Gamma$ which satisfy the Poisson bracket relations 
\be 
  \{C_I,C_J\}\ =\ a_{IJ}^KC_K \ ,
\ee
with the coefficients $a_{IJ}^K$ being constant numbers. Then, the constraints define a $d$-dimensional Lie group, say $G$,  of the gauge
transformations of the theory. The right action of the group will be denoted by
\be 
  G\times \Gamma\ni (g,\gamma)\ \mapsto\ \gamma g \ .
\ee
Every constraint function $C_I$ corresponds to a left invariant vector field $\xi_I$ tangent to $G$ such that  for every 
function $f:\Gamma\rightarrow \mathbb{R}$,
\be 
  \frac{\rd}{\rd t}f(\gamma \exp(t\xi_I))\ =\ \{f,C_I\}(\gamma) \ .
\ee

A (strong) Dirac observable of that theory, is every function
$F:\Gamma\rightarrow \mathbb{R}$ invariant with respect to the
action of the gauge group $G$. That definition is complete, but from the point
of view of the applications in the quantum theory, it is important
to have an analytic formula that expresses a given Dirac observable
by some explicitly known functions on $\Gamma$ and their Poisson
brackets. Such observables are provided by the
framework of the relational observables \cite{rel-obs}. We introduce now
our generalized formulation of this framework motivated by \cite{Marolf2}
(in the main part of our paper which concerns a $1$-constraint case,
our formula anyway reduces to that of \cite{Marolf2}).

Given: a function $F:\Gamma\rightarrow\mathbb{R}$, a point $\gamma\in\Gamma$, and
$g\in G$, we will denote by $F(\gamma\cdot )$, and, respectively, $F(\cdot g)$ the
following functions
\be 
  F(\gamma\cdot):G\ni g\ \mapsto\ F(\gamma g)\ ,\quad {\rm and}\quad 
  F(\cdot g):\Gamma\ni \gamma\ \mapsto\ F(\gamma g) \ .
\ee
To turn functions defined on $\Gamma$ into Dirac observables, we choose
sufficiently generic reference functions $T^I:\Gamma\rightarrow
\mathbb{R}$, $I=1,...,d$. Ideally, each set of points defined by
condition
\be 
  T^1\ =\ t^1,\quad \ldots,\quad T^d\ =\ t^d,\qquad t^1,\ldots,t^d\in \mathbb{R}
\ee
should define codimension $d$ submanifold in $\Gamma$ transversal to
the orbits of the group $G$ and intersecting each orbit in at most
one point. This  conditions can be relaxed, by assuming it holds on
a sufficiently small neighborhood of a given point
$\gamma_0\in\Gamma$. Then, given:
\begin{itemize}
  \item a function $F:\Gamma\rightarrow \mathbb{R}$ --the function we
    want to ``observe''-- of a support in the neighborhood of $\gamma_0$,
  \item a point $\gamma\in \Gamma$, and 
  \item numbers $t^1,\ldots,t^d\in \mathbb{R}$,
\end{itemize}
the function $F^{D}_{(T^1,\ldots,T^d,t^1,\ldots.t^d)}$ defined via the integral
\begin{equation} \label{eq:obs-int1}
  \int_G |\rd T^1(\gamma\cdot)\wedge \ldots \wedge
  \rd T^d(\gamma\cdot)\,|\, F(\gamma\cdot) \prod_{I=1}^d
  \delta(T^I(\gamma\cdot)-t^I)\ =:\ F^{D}_{(T^1,\ldots,T^d,t^1,\ldots,t^d)}(\gamma) \ .
\end{equation}
is a Dirac observable, whenever well defined.\footnote{The restriction on the 
  support of $f$ can be passed in the standard way to a partition of unity, 
  a family of functions $(\kappa_A)_A$, whose supports satisfy the suitable 
  condition, and such that $\sum_A\kappa_A\ =\ 1.$ However, still the integral 
  may take infinite value or be undefined  at a given point $\gamma_0$, for 
  example when both the group $G$ and the orbit passing through $\gamma$, 
  respectively, is one dimensional, but there are infinitely many different 
  elements $g_1,...,g_n,...\in G$ which act on $\gamma$ as identity.}
On the intuitive level, $F^{D}_{(T^1,\ldots,T^d,t_1,\ldots,t^d)}(\gamma)$ is  $F(\gamma')$ where
$\gamma'$ is the intersection of the gauge group orbit passing through $\gamma$ with
the surface $T^I=t^I$.

The formula will be even more useful when we express it
in terms of a left invariant  Haar measure on $G$. For this purpose
we use a decomposition of each  1-form  $\rd T^I(\gamma\cdot)$ on $G$
into a coframe of left invariant 1-forms $(\omega^1,\ldots,\omega^d)$,
dual to a frame $(\xi_1,\ldots,\xi_d)$ of left invariant vector fields
on $G$, namely
\be 
  \rd T^I(\gamma\cdot)|_{\cdot =g}\ =\ \xi_J(T^I(\gamma\cdot)|_{\cdot=g}\omega^J\ 
  =\ \{T^I(\cdot g),C_J\}|_{\cdot =\gamma}\omega^I \ .
\ee
The coframe determines a normalization of the left invariant Haar measure
\begin{equation}
  \rd\mu_{\rm H}^{\rm L}\ =\ |\omega^1\wedge\,...\,\wedge\omega^d|\ .
\end{equation}
The integral \eqref{eq:obs-int1} can now be expressed in the following form
\be 
  F^{D}_{(T^1,\ldots,T^d,t^1,\ldots,t^d)}(\gamma)\ =\ \int_G\rd\mu_{\rm H}^{\rm L}(g)
  \,\frac{1}{n!}\,|\,
  \epsilon^{I_1\ldots I_d}\{T^1,C_{I_1}\}(\gamma g)\ldots\{T^d, C_{I_d}\}(\gamma g)
  \,|\,
  F(\gamma g) \prod_{I=1}^d \delta(T^I(\gamma g)-t^I) \ , \label{classobs}
\ee
which will be used in the next subsection as the basis for quantization.

\subsection{Quantization}\label{GAq} 

A general scheme  of the corresponding quantum theory is not complete. 
Some technical details can be fixed only
when we pass to examples (it is conceivable in some pathological cases the scheme does
not work). We will however successfully complete it for the discrete and continuous case
studied in Section \ref{Exp1} and \ref{Exp2}.

Let ${\cal H}_{\rm kin}$ be a Hilbert space in which to every function 
$F:\Gamma\rightarrow \mathbb{R}$ we assign an operator $\hat{F}$ defined modulo 
the ordering ambiguity, and such that the known quantization relations are satisfied. 
In particular, let $\hat{C}_1,\,ldots\, ,\hat{C}_d$ be quantum constraint operators and
suppose
\begin{equation}
  G\ni g\ \mapsto\ U(g)\in U({\cal H}_{\rm kin})
\end{equation}
is the corresponding unitary representation of the group $G$. (We are assuming here 
the quantum constraints generate a group of unitary transformations in the Hilbert 
space ${\cal H}_{\rm kin}$ isomorphic to $G$.)

Our aim now is to define a quantum counterpart of the classical constraint equation
$C_I=0$, and the quantum Dirac observables, quantum counterparts of (\ref{classobs}).

Suppose now there exists a decomposition of ${\cal H}_{\rm kin}$ into 
irreducible unitary representations of $G$,
\be 
  {\cal H}_{\rm kin}\ \equiv \int^\oplus \rd\mu(\rho){\cal H}_{{\rm kin},\rho} \ ,
  \label{irrdecomp}
\ee
where throughout this paper we use the notion of a formal integral of Hilbert spaces
defined for a measurable space $(X,\mu)$, and a family of the Hilbert spaces
$({\cal H}_x)_{x\in X}$ equipped with the Hilbert space structure
\begin{subequations}\begin{align} 
  (v_x)_{x\in X}\ +\alpha(w_x)_{x\in X}\ &:= (v_x+\alpha w_x)_{x\in X} \ , \\
  ((v_x)_{x\in X}\,|\,(w_x)_{x\in X})\ &:=\ \int \rd\mu(x) (v_x|w_x)_x \ .
\end{align}\end{subequations}
The formal integral Hilbert space and its elements, respectively, will be denoted by
\begin{equation}
  \int^\oplus_X\rd\mu(x){\cal H}_x\ \ni\ \int^\oplus_X\rd\mu(x)v_x \ .
\end{equation}
Here, the measurable set $X$ is the space of the irreducible representations of $G$, 
and each of the Hilbert spaces ${\cal H}_{{\rm kin},\rho}$ has the structure
\be 
  {\cal H}_{{\rm kin},\rho} \ =\ V_\rho\otimes \tilde{\cal H}_{{\rm kin},\rho} \ ,
\ee
where the action of the group $G$ in ${\cal H}_{{\rm kin}}$ passes to
\be 
  U(g)(v_\rho\otimes\tilde{\psi}_\rho)_{\rho\in X}\ 
  =\ (\rho(g)v_\rho\otimes\tilde{\psi}_\rho)_{\rho\in X} \ .
\ee

To define solutions to the quantum constraints $\hat{C}_I$, $I=1,\ldots,d$ we need 
some extra structure of continuity around the trivial representation $\rho_0$, which makes a limit
\be 
  \lim_{\rho\rightarrow \rho_0}{\cal H}_{{\rm kin},\rho} \label{H_0}
\ee
well defined. For example, this is the case if there is a natural isomorphism 
$\tilde{\cal H}_{{\rm kin},\rho}\equiv \tilde{\cal H}_{{\rm kin},\rho_0}$ for $\rho$ 
sufficiently close to $\rho_0$. Such a situation may occur in case where all constraints commute.
As usually ``$\equiv$'' means the existence of a unitary isomorphism between the Hilbert spaces.
The methods for singling out ${\cal H}_{{\rm kin}\rho_0}$ are presented in \cite{Maurin}. 
We briefly sketch them in Appendix \ref{ma_ap}.

Having said all that, we define:
\begin{df} A solution to a quantum constraint defined by the constraint operators 
  $\hat{C}^1,\, \ldots\, ,\hat{C}^d$ is each element $\psi$ of the Hilbert space 
  ${\cal H}_{{\rm kin},\rho_0}$ in the decomposition (\ref{irrdecomp}). A Hilbert space 
  of solutions is the Hilbert space  ${\cal H}_{{\rm kin},\rho_0}$.
\end{df}
Each solution $\psi\in {\cal H}_{{\rm kin},\rho_0}$ to the constraints can be thought of as a functional
\be 
  {\cal H}\ni\psi'\ \mapsto\ (\psi\,|\,\psi'_{\rho_0})_{\rho_0} \label{funct}
\ee
well defined on the domain of elements of ${\cal H}_{\rm kin}$ represented by families 
$(\psi'_\rho)_\rho$  {continuous in $\rho$ (this is where the notion of the continuity is relevant).}

Having the physical Hilbert space defined, lets turn our attention to the observables.
A Dirac observable $\hat{F}^{\rm D}$ is an operator in ${\cal H}$
corresponding to a classical Dirac observable $F^{\rm D}$ and
compatible with the decomposition (\ref{irrdecomp}) in the natural
way
\be 
  \hat{F}^{\rm D}(\psi_\rho)_\rho\ =\ (\hat{F}^{\rm D}_\rho\psi_\rho)_\rho \ .\label{Frho}
\ee
We will assume (and prove in he cases considered below and introduced in Section 
\ref{sec:intro}) some continuity of $\hat{F}^{\rm D}(\psi_\rho)_\rho$ with respect 
to $\rho$ on the vectors from the domain ${\mathcal D}$. A formal definition will be provided 
in appendix \ref{ma_ap}. The Dirac observable defined 
in the Hilbert space of solutions is the component operator $\hat{F}^{\rm D}_{\rho_0}$ 
acting in the physical Hilbert space ${\cal H}_{\rho_0}$. An equivalent definition of 
this operator in terms of (\ref{funct}), is provided just by the duality.

In the class of examples considered in this paper, the quantized
version of the  integral (\ref{classobs}), that is
\be 
  \hat{F}^{D}_{(T^1,...,T^d,t^1,...,t^d)}(\gamma)\ 
  =\ \frac{1}{i^d}\int_G\rd \mu_{\rm H}^{\rm L}(g)\,
  U(g)^{-1}\,{\rm Sym}\left(\,\frac{1}{n!}|\,
  \epsilon^{I_1\ldots I_d}[\hat{T}^1,\hat{C}_{I_1}]\ldots [\hat{T}^d, \hat{C}_{I_d}]
  \,|\,
  \hat{F} \prod_{I=1}^d \delta(\hat{T}^I-t^I\hat{\id})\right)\,U(g) \label{qobs}
\ee
is well defined operator in the kinematical Hilbert space ${\cal H}$ and takes the form 
(\ref{Frho}), therefore it defines an observable in the Hilbert space of solution 
${\cal H}_{\rho_0}$. The symbol ``Sym'' stands for a symmetrization making the product of
non-commuting operators a symmetric operator (or, at least a symmetric sesquilinear form 
on the domain $\mathcal D$). Explicit form of this symmetrization will be adjusted to specific 
examples in order to recover correct results for some testing model observables.
\medskip

\noindent{\bf Remark} If  the constraint functions in the classical
theory are defined modulo transformations
\be 
  {C'}_I\ =\ {N}_I^J C_J 
\ee
and $N_I^J$ are functions on $\Gamma$ themselves, then the quantum
counterpart may lead to several ambiguities: factor ordering, 
the particular form of the resulting space of solutions. 
We will discuss them in context of particular examples studied in the next 
subsection.

\subsection{$1$-constraint Hamiltonian systems}\label{GAFRW} 

The structure of a constrained Hamiltonian system considered in this article stems 
from the Loop Quantum Cosmology 
models of isotropic and homogeneous universes (of negative
in the first example, and positive in the second one cosmological constant) filled 
with a massless scalar field \cite{APS-imp,bp-negL,ap-posL}.
In the paper we will introduce only those general elements and assume only 
those general properties which  are needed for our characterization of the solutions 
to the quantum constraint and for the definition of the Dirac observables. In 
this sense the presented constructions will be somehow abstract, but directly applicable
to the models they stem from.
Although, our investigation regards the quantum theory, we invoke first some 
information about the classical one in order to provide an intuition needed as a 
basis for the constructions implemented in the quantum models.

The kinematical phase space $\Gamma_{\rm kin}$ is the Cartesian
product $\Gamma_{\rm sc}\times \Gamma_{\rm gr}$, where 
$\Gamma_{\rm sc}=\{(T,\Pi)\in \mathbb{R}^2\}$ ($T$ is the homogeneous scalar field 
and $\Pi$ its momentum), with the Poisson bracket
\be 
  \{F,G\}_{\rm sc}\ =\ \frac{\partial F}{\partial T}\frac{\partial G}{\partial\Pi}
  -\frac{\partial F}{\partial \Pi}\frac{\partial G}{\partial T} \ ,
\ee
and $\Gamma_{\rm gr}=\{(v,p_v)\in \mathbb{R}^2\}$ is the part corresponding 
to the gravitational degrees of freedom.

The gauge group $G$  is 1-dimensional, isomorphic to $\mathbb{R}$. The left invariant 
vector fields tangent to it correspond to  one constraint
\be 
  {C}\ =\ \frac{1}{2}{\Pi}^2 B \ +\ {C}_{\rm gr}\label{classC} \ ,
\ee
where $B$ and $C_{\rm gr}$ are functions defined on the gravitational phase space 
$\Gamma_{\rm gr}$. The function $C$ is at the same time the Hamiltonian of the 
theory, that is it generates the physical evolution of the system.

The system admits one obvious Dirac observable: the function $\Pi$. To
construct the other ones we apply the method described in Section \ref{GAclass}
and specified via an integral (\ref{qobs}). As a time variable we choose the 
function $T$.

For the 1-dimensional gauge group and upon the choices specified above, the integral (\ref{qobs})
corresponding to a given kinematical observable $F$ (a function defined on
$\Gamma_{\rm sc}\times\Gamma_{\rm gr}$) reads
\be 
  {F}^{D}_{(T,t)}(\gamma)\ =\ \int \rd\tau \,|\, \Pi(\gamma(\tau))B(\gamma(\tau))
 \,|\, {F}(\gamma(\tau)) \delta({T}(\gamma(\tau))-t) \ , \label{qobs1}
\ee
where
\be 
  \tau\mapsto\gamma(\tau) \ ,\qquad \gamma(0)\ =\ \gamma
\ee
is the flow generated by $C$, and the Poisson bracket in
(\ref{classobs}) was replaced via use of the identity,
\begin{equation} 
  \{T,C\}\ =\ \Pi B \ .
\end{equation}

The question relevant for our studies is what is the form and properties of
$F^D_{(T,t)}$, once as $F$ we select a Dirac observable, that is if
\be 
  \{F,C\}\ =\ 0 \ . 
\ee
The answer to it depends on the properties of the flow generated in the
phase space by the Hamiltonian constraint $C$.

If  the function $T$ restricted to every orbit of the flow of $C$,
ranges from $\mp \infty$ to $\pm \infty$, and the map $G\ni g\mapsto \gamma g$ is 1-1, then
\be 
  F^D_{(T,t)}(\gamma)\ =\ F(\gamma)\quad \text{whenever}\quad \Pi(\gamma)\neq 0 \ .
\ee
On the other hand
\be
  \Pi(\gamma_0)\ =\ 0 \ , \ \text{and}\ T(\gamma_0)\ \neq t\quad
  \text{implies}\quad F^D_{(T,t)}(\gamma_0)\ =\ 0 \ .\label{Pi=0}
\ee
For this case an application of the construction \eqref{qobs} to $\Pi$ itself
gives the function itself
\be 
  \Pi^D_{(T,t)}\ =\ \Pi \ . 
\ee 
An example of a classical model of that property is provided by the discrete 
case defined in Section \ref{sec:GA-intro} and studied in detail in Sections 
\ref{sec:res} and \ref{Exp1}.

Another possibility, represented by the continuous case studied in 
Section~\ref{Exp2}, is that $T$ restricted to each orbit of the flow of $C$ is 
bounded, and its supremum/minimum $T^\pm(\gamma)$ depends on a point $\gamma$. 
Then, the function $F^D_{T,t}(\gamma)$ corresponding to given Dirac observable $F$ 
and chosen value of the parameter $t$ takes the form
\begin{equation} 
  F^D_{(T,t)}(\gamma)\ =\ \begin{cases}
    F(\gamma) \ ,  & \text{if } t\in [T^-(\gamma),T^+(\gamma)] \ ,\\
    0 \ ,  &  \text{otherwise.}
  \end{cases} 
\end{equation}
The above formula is a generalization of \eqref{Pi=0} which thus can be thought 
of as just a special case of it.

Having at our disposal the above classical framework, which is adopted to the $1$-constraint
case studied in this subsection, we can now turn our attention to its quantum counterpart.
We construct it by restricting the prescription presented in Section \ref{GAq} in a 
way analogous to the one performed above on the classical level.

For considered class of systems the kinematical Hilbert space takes the following form,
\be 
  {\cal H}_{\rm kin}\ =\ L^2(\mathbb{R})\otimes {\cal H}_{\rm gr} \ ,
\label{Hkin}
\ee
with ${\cal H}_{\rm gr}$ being some general Hilbert space whose detailed 
properties are not relevant for our studies in the discrete case, and they will be 
further specified in the continuous case. The scalar products in the Hilbert spaces 
${\cal H}_{\rm kin}$, and ${\cal H}_{\rm gr}$, respectively, will be denoted by 
$(\cdot|\cdot)_{\rm kin}$, and, respectively, $(\cdot|\cdot)_{\rm gr}$.
The operators $\hat{T}$ and $\hat{\Pi}$ are defined in
$L^2(\mathbb{R})$ as
\be 
  \hat{T}\psi(T)\ =\ T\psi(T) \ ,\qquad 
  \hat{\Pi}\psi(T)\ =\ \frac{1}{i}\frac{\partial}{\partial T}\psi(T) \ . 
\ee

The quantized scalar constraint takes the following form
\be 
  \hat{C}\ =\ \frac{1}{2}\hat{\Pi}^2\otimes \hat{B} \ +\ \hat{\id}\otimes\hat{C}_{\rm gr} \ ,
  \label{C}
\ee
where the operators $\hat{B}$, $\hat{B}^{-1}$ and $\hat{C}_{\rm gr}$ are
defined in a same domain ${\cal D}_{gr}\subset{\cal H}_{\rm gr}$, each of
them is essentially self adjoint. We will be also assuming that $\hat{B}$ is bounded
($\hat{B}^{-1}$  may be unbounded).

The decomposition (\ref{irrdecomp}) needed for the identification of the space of solutions
is provided just by spectral decomposition of the operator $\hat{C}$, that is
\be 
  {\cal H}_{\rm kin}\ \equiv\ \int^{\oplus} \rd\mu_{\hat{C}}(c){\cal H}_{{\rm kin},c} \ .
\ee
where ${\cal H}_{{\rm kin},c}$ is a family of the Hilbert spaces, labeled by $c\in\mathbb{R}$
and having a natural vector space structure and the Hilbert product
\begin{equation}
  ((\psi_c)_c|(\psi_{c'})_{c'})\ =\ \int \rd\mu_{\hat{C}}(c)(\psi_c|\psi'_c)_c \ .
\end{equation}
with the action of the operator $\hat{C}$ being
\be 
  \hat{C}(\psi_c)_c\ =\ (c\psi_c)_c \ .
\ee

For the constraint of the form \eqref{C}, to find the spectral decomposition 
of the operator $\hat{C}$, it is easier to first use the spectral decomposition 
defined by the operator $\hat{\Pi}$
\be 
  {\cal H}_{\rm kin}\ =\ \int^\oplus \rd\mu_{\hat{\Pi}}(\ppi){\cal
  H}_{\rm kin,\ppi} \ , \label{Pidecomp}
\ee
as it is quite explicit,
\be 
  \rd\mu_{\hat{\Pi}}(\ppi)\ =\ \rd\ppi \ , \qquad 
  {\cal H}_{{\rm kin},\ppi}\ =\ {\cal H}_{\rm gr} \ .
\ee
The unitary map  $L^2(\mathbb{R})\otimes{\cal H}_{\rm gr}\ni \psi\
\mapsto\ \int^\oplus \rd\ppi\psi_\ppi$ is in this case defined by
\be 
  \psi(T)\ =\ \int_{-\infty}^\infty \frac{\rd\ppi}{\sqrt{2\pi}}
  e^{i\ppi T}\psi_\ppi \ , \label{psifurier}
\ee
where $\psi$ is thought of as a ${\cal H}_{\rm gr}$ valued function
$T\mapsto\psi(T)$.

In the next step we apply the above decomposition, considering for each 
$\ppi\in\mathbb{R}$ the operator
\be 
  \hat{C}_\ppi\ := \frac{1}{2}\ppi^2 \hat{B}\ +\ {\hat{C}_{\rm gr}} \ , \label{Cpi} 
\ee
defined in the domain ${\cal D}_{gr}\subset {\cal H}_{\rm gr}$, and the spectral 
decomposition corresponding to it. As a result we arrive to a joint 
spectral decomposition
\be 
  {\cal H}_{\rm kin}\ =\ \int^\oplus \rd\ppi \rd\mu_{\hat{C}_\ppi}(c){\cal
  H}_{{\rm kin},\ppi c} \ . \label{Hpic} 
\ee
As the value of $\ppi$ enters explicitly into the measure $\rd\mu_{\hat{C}_\ppi}(c)$, the 
above construction is sensitive to the order in which it was performed: $\hat{\Pi}$
first, $\hat{C}$ second. In consequence one cannot immediately extract the Hilbert 
structure corresponding to the space of the solutions to the constraint. However, 
the joint spectrum endowed with a measure is independent of the order. To rewrite the 
decomposition into a useful form we need to invert the order, by reexpressing 
the measure in (\ref{Hpic}) as
\be 
  \rd\ppi \rd\mu_{\hat{C}_\ppi}(c)\ =\ \rd\mu'(c)\rd\mu'_c(\ppi)\ =:
  \rd\mu_{\hat{C}}(c)\rd\mu_c(\ppi) \ ,
\ee
and constructing the desired decomposition
\be 
  {\cal H}_{\rm kin}\ =\ \int^\oplus \rd\mu_{\hat{C}}(c)\int^\oplus
  \rd\mu_c(\ppi){\cal H}_{{\rm kin},\ppi c} \ .\label{Hcpi}   
\ee
On the technical level our aim is the characterization of the structure of 
the Hilbert spaces 
\begin{equation} 
  {\cal H}_{\rm kin,c}=\int^\oplus \rd\mu_c(\ppi){\cal H}_{{\rm kin},\ppi c}
\end{equation}
in a neighborhood of $c=0$, as well as the structure of its elements.

Once we have the Hilbert space structure of the space of solutions to the constraint
at our disposal, the next task is to define useful Dirac observables, by providing 
a precise meaning to the formula (\ref{qobs}). In the case at hand that integral reads
\be 
  \hat{F}^{D}_{(T,t)}\ =\ \int \rd\tau e^{-i\tau \hat{C}} {\rm Sym}
  \left(\,|\,\hat{\Pi}\hat{B}
  \,|\, \hat{F} \delta(\hat{T}-t\hat{\id})\right)e^{i\tau\hat{C}} \ , \label{qobsC}
\ee
and is identified with the following sesquilinear form defined in
${\cal H}_{\rm kin}$
\be
  (\psi,\psi')\ \mapsto\ (\psi\,|\, \hat{F}^{D}_{(T,t)}\psi')\ :=\
  \int \rd T \int \rd\tau \left((e^{i\tau \hat{C}}\psi)(T)\,|\, \left({\rm Sym} 
  \left(\,|\,\hat{\Pi}\hat{B}
  \,|\, \hat{F} \delta(\hat{T}-t\hat{\id})\right)(e^{i\tau\hat{C}}\psi')\right)(T)\right)_{\rm gr} \ .
  \label{psiqobspsi} 
\ee
In the examples studied below, that form will be defined by an operator
preserving the spectral decomposition of the operator $\hat{C}$ and
defining by the duality an operator in each subspace ${\cal H}_{{\rm
kin},c}$, in particular at $c=0$.

It is worth remembering, that the operator $\hat{\Pi}$ is a quantum Dirac 
observable itself, without using the integral. However, we will also construct 
$\hat{\Pi}^{\rm D}_{(T,t)}$ for the comparison.

\section{The results}\label{sec:res}

For the rest of this paper we focus on the detailed studies of the systems with the 
structure of the constraints as defined in Section \ref{GAFRW}. For the clarity 
of the presentation,  we provide in this section an outline of the results as well as 
the detailed sketch of the techniques used to arrive to them. The detailed proofs and 
derivations are presented in the next two sections, corresponding, respectively, to the 
discrete and continuous case, as defined at the end of Section \ref{sec:intro}.

The departure point for the rest of this paper is the spectral/GA framework  
(\ref{Hkin}--\ref{psiqobspsi}). We examine the two cases mentioned above and contrast
the differences. The discrete case turns out to be a model case in which the Hilbert 
space of solutions to the constraint and the Dirac observables derived by using the spectral/GA framework (\ref{Hkin}--\ref{psiqobspsi}) coincide with those following from the Schr\"odinger picture (\ref{3a}).
In the continuous case, on the other hand, the result of (\ref{Hkin}--\ref{psiqobspsi}) 
is different, than the one obtained from (\ref{3a}) and the properties of the observables 
are even more surprising.

Both the considered cases are defined just by a set of assumptions outlined below:
\begin{enumerate}[I.]
  \item{\bf Discrete case} (Section \ref{Exp1}) 
    For every $\ppi\in\mathbb{R}$, the operator 
    $\hat{C}_\ppi=\frac{1}{2}\ppi^2\hat{B} + \hat{C}_{\rm gr}$ is assumed to define 
    an orthonormal basis $\{e_{\ppi,c_n(\ppi)}\,|\,n\in\mathbb{N}\}$ of the Hilbert 
    space ${\cal H}_{\rm gr}$ which consists of eigenvectors,
    \begin{equation}
      \hat{C}_\ppi e_{\ppi,c_n(\ppi)}\ =\  c_n(\ppi)e_{\ppi,c_n(\ppi)} \ .
    \end{equation}
    Additional technical assumptions  ensure the  continuity, differentiability and 
    non-degeneracy  of relevant functions constructed from the map 
    $(\ppi,n)\mapsto (c_n(\ppi),e_{\ppi,c_n(\ppi)})$ (of course the map is even 
    in the variable $\ppi$). We are also assuming that
    \begin{equation} 
      \hat{C}_{\rm gr}\ \leq\ -c_0\ <\ 0 \ .
    \end{equation} \label{it:discr-ass}
  \item{\bf Continuous case} (Section \ref{Exp2}) 
    The Hilbert space  ${\cal H}_{\rm gr}$ is further specified as $L^2(\mathbb{R},d\nu_0)$ 
    with a suitable measure. The domain on which each of the  operators 
    $\hat{C}_\ppi$ including ${\hat C}_{\rm gr}$ is essentially self adjoint is the 
    subspace ${\cal D}_{gr}$ of $C^\infty_0(\mathbb{R})$ (smooth functions of the compact support)\footnote{In 
      the LQC example which gave rise to this case, the measure 
      $d\nu_0(v)=\sum_{n\in\mathbb{N}}\delta(v-n)$, and the differentiability does not 
      play a role}. The operators $\hat{B}$ and ${\hat C}_{\rm gr}$ are extended by 
    the duality onto the space of functions on $\mathbb{R}$ dual to $C^\infty_0$.  
    For every $\ppi\ge 0$, there is a normalized to the Dirac delta basis  
    $\{e_{\ppi,c}\,|\,c\in\mathbb{R}\}$ of ${\cal H}_{\rm gr}$ which consist of the 
    eigenfunctions of $\hat{C}_\ppi$,
    \begin{equation}
      \hat{C}_\ppi e_{\ppi,c}\ =\  c\,e_{\ppi,c} \ .
    \end{equation}
    The key assumption which makes this case essentially different\footnote{The obvious 
      difference is the continuity of the spectrum of $\hat{C}_{\ppi}$, but it would not 
      be sufficient for the peculiar properties that emerge in that case.} than the 
    previous one, concerns the asymptotic behaviour of the eigenfunctions, namely
    we impose
    \begin{equation}
      \lim_{V\rightarrow \infty}\int_{-V}^V \rd\nu_0(v) 
      \left( \overline{e_{\ppi,c}(v)}\hat{C}_{\rm gr}e_{\ppi,c'}(v)\ 
      -\  \overline{\hat{C}_{\rm gr}e_{\ppi,c}(v)}e_{\ppi,c'}(v)\right)\ 
      =\ b\sin(a(\ppi,c)-a(\ppi,c')) \ ,
    \end{equation}
    where $b\in \mathbb{R}$ is a constant and $(\ppi,c)\mapsto a(\ppi,c)$ is a function.
    Again, extra technical assumptions are imposed to ensure the non-degeneracy and 
    the differentiability of the relevant structures. \label{it:case-cont}
\end{enumerate}

For each of the two cases defined above we construct the physical Hilbert space 
and Dirac observables by applying the methods specified in Section \ref{GAFRW}. This is 
performed in the following sequence of steps:

$\bullet$ 
The first step  is {\bf finding the Hilbert spaces ${\cal H}_{{\rm kin},c}$}, components of the
righthand side of \eqref{Hpic} by identifying the complete spectral decomposition
of the quantum constraint operator
$\hat{C}=\frac{1}{2}\hat{\ppi}^2\otimes\hat{B}+1\otimes{\hat C}_{\rm gr}$.
Since in both the discrete and the continuous case $c=0$ is a measure zero point 
of the spectrum, we first consider solutions to a quantum constraint $\hat{C}-c$ 
for arbitrarily fixed value $c\in\mathbb{R}$ in a neighborhood of $c=0$, before
setting $c=0$. For both cases we define a set of (auxiliary) Hilbert spaces ${\cal H}_{{\rm kin},c}$ 
which are formed by the functions (``solutions'' to the constraint)
\begin{equation}
  E_{\ppi,c}\ :\ T\ \mapsto\ \frac{1}{\sqrt{2\pi}}e^{i\ppi T}e_{\ppi,c},\label{Tmapstoe} \ . 
\end{equation}
The set of the values $\ppi$ and the scalar product depend on the case.

In the continuous case, $\ppi$ runs through the set of all the real numbers, and 
the scalar product in ${\cal H}_{{\rm kin},c}$ (that is the ``physical'' scalar product 
of ${\cal H}_{{\rm kin},c}$) is
\begin{equation}
  (E_{\ppi,c}\,|\,E_{\ppi',c})_{{\rm kin},c}\ =\  \delta(\ppi-\ppi') \ . \label{sprII}
\end{equation}

In the discrete case, the value of $\ppi$ in $E_{\ppi,c}$ ranges a discrete set $\{\pm\ppi_n(c)\,:\, n\in\mathbb{N}\}$ depending on the fixed value of $c$,
where, given $n$, the function  $c\mapsto \ppi_n(c)$ is the inverse function to
$\ppi\mapsto c_n(\ppi)$ restricted to  $\ppi\ge 0$. Notice, that
\be 
  \ppi_n(c)\ >\ 0 \ ,\quad \text{for every}\quad n\in\mathbb{N}
\ee
due to the assumption $c\in (-c_0,c_0)$.

The scalar product in ${\cal H}_{{\rm kin},c}$ equals in that case
\begin{subequations}\begin{align}
  (E_{\ppi_n(c),c}\,|\,E_{\ppi_{n'}(c),c})_{{\rm kin},c}\
  &=\  \frac{dc_n}{d\ppi}|_{\ppi=\ppi_n(c)} \delta_{n,n'}\ 
   = (E_{-\ppi_n(c),c}\,|\,E_{-\ppi_{n'}(c),c})_{{\rm kin},c} \ ,   \label{EEI}\\
  (E_{-\ppi_n(c),c}\,|\,E_{\ppi_{n'}(c),c})_{{\rm kin},c}\ &=\ 0 \ .   \label{EE-I}
\end{align}\end{subequations}

$\bullet$
The second step, is to view the quantum constraint and its solutions (\ref{Tmapstoe}) as a {\bf unitary evolution} in a suitable Hilbert space ${\cal H}_c$ formed by the eigenvectors/eigenfunctions $e_{\ppi,c}$ and equipped with a new  scalar product $(\cdot\,|\,\cdot)_c$ which replaces the old one $(\cdot|\cdot)_{\rm gr}$. The new scalar product $(\cdot|\cdot)_c$ is determined by (\ref{EEI},\ref{EE-I}) and, respectively (\ref{sprII}), to be
\begin{subequations}\begin{align}
  (e_{\ppi_n(c),c}\,|\,e_{\ppi_{n'}(c),c})_c\ 
  &=\ 2\pi (\frac{d\ppi_n(c)}{dc})^{-1}\delta_{n,n'} \ , & 
  &\text{in the discrete case,} \\
  (e_{\ppi,c}\,|\,e_{\ppi',c})_c\ 
  &=\ \delta(\pi-\pi') \ , & 
  &\text{in the continuous case,} \label{sprcII}
\end{align}\end{subequations}
where $\ppi\in \mathbb{R}^+$, because $e_{-\ppi,c}=e_{\ppi,c}$.

With the scalar product $(\cdot\,|\,\cdot)_c$, the product $(\cdot\,|\,\cdot)_{{\rm kin},c}$ between the solutions to the quantum constraint  can be evaluated at any ``instant'' of the variable $T$ in $E_{\ppi,c}(T)\ =\ \frac{1}{\sqrt{2\pi}}e^{i\ppi T}e_{\ppi,c}$. In the discrete case, the equalities (\ref{EEI},\ref{EE-I}) give
\begin{subequations}\begin{align}
  (E_{\ppi_n(c),c}\,|\,E_{\ppi_{n'}(c),c})_{{\rm kin},c}\ &=\ (E_{\ppi_n(c),c}(T)\,|\,E_{\ppi_{n'}(c),c}(T))_c \ ,\\
  (E_{-\ppi_n(c),c}\,|\,E_{-\ppi_{n'}(c),c})_{{\rm kin},c}\ &=\ (E_{-\ppi_n(c),c}(T)\,|\,E_{-\ppi_{n'}(c),c}(T))_c \ ,\\
  (E_{-\ppi_n(c),c}\,|\,E_{\ppi_{n'}(c),c})_{{\rm kin},c}\ &=\ 0 \ .
\end{align}\end{subequations}
whereas in the continuous one due to (\ref{sprII}) the scalar product reads
\begin{subequations}\begin{align}
  (E_{\ppi,c}\,|\,E_{\ppi',c})_{{\rm kin},c}\
  &=\  (E_{\ppi,c}(T)\,|\,E_{\ppi',c}(T))_c \ , & 
  &\text{for } \pi\pi' \ge 0 \ ,\\
  (E_{\ppi,c}\,|\,E_{\ppi',c})_{{\rm kin},c}\
  &=\ 0 \ , &  
  & \text{otherwise.}
\end{align}\end{subequations}
Having at our disposal the scalar product we can define the unitary evolution. 
To do so we note that a map
\begin{equation} 
  (e_{\ppi,c},e_{\ppi',c})\ \mapsto\ E_{|\ppi|,c}+E_{-|\ppi|,c}
\end{equation}
determines a unitary Hilbert space isomorphism
\be 
  {\cal H}_c\oplus {\cal H}_c\ \rightarrow\  {\cal H}_{{\rm kin},c} \ .\label{HcHc}
\ee
By ${\cal H}^+_{{\rm kin},c}$ and, respectively, ${\cal H}^-_{{\rm kin},c}$
we denote the images of the first, and respectively the second term.

Finally, the unitary evolution
\begin{equation} 
  U(T)\ :\ {\cal H}_c\oplus {\cal H}_c\ \rightarrow\ {\cal H}_c\oplus {\cal H}_c
\end{equation}
dictated by the quantum constraint $\hat{C}-c\hat{\id}$ amounts to
\begin{equation} 
  (e_{\ppi,c},e_{\ppi',c})\ \mapsto\ (E_{|\ppi|,c}(T), E_{-|\ppi'|,c}(T)) \ .
\end{equation}

Note that up to this point the only difference between the discrete and 
the continuous case is in the discreteness versus the continuity of the label $\ppi$.

$\bullet$
The third step, is to find a {\bf relation of the scalar product} 
$(\cdot\,|\,\cdot)_c$ and of the {\bf Hilbert space} ${\cal H}_c$ {\bf with} 
${\cal H}_{\rm gr}$ and its scalar product. This step is both important and 
nontrivial, because the eigenvectors/eigenfunctions $e_{\ppi,c}$ have been defined 
in terms of ${\cal H}_{\rm gr}$, and given fixed $\ppi$, the corresponding set of 
$e_{\ppi,c}$s is a basis of ${\cal H}_{\rm gr}$, orthonormal in the discrete case, 
and, respectively, Dirac delta-orthonormal in the continuous one. However, 
now we fix $c$ and let $\ppi$ be arbitrary. The product $(\cdot\,|\,\cdot)_c$ has 
been introduced just by declaring its values $(e_{\ppi,c}|e_{\ppi',c})_{c}$. A clue 
in how to relate these two inner products is provided by the following equation 
satisfied by the eigenfunctions/eigenvectors $e_{\ppi,c}$,
\be
  \label{clue} 2\hat{B}^{-1}(\hat{C}_{\rm gr}-c\hat{\id})e_{\ppi,c}\ =\ -\ppi^2e_{\ppi,c} \ . 
\ee
From the symmetry of ${\hat{C}_{\rm gr}}$ follows immediately, that the operator on 
the left hand side is symmetric in the domain ${\cal D}_{gr}$ of the operator ${\hat{C}_{\rm gr}}$ with 
respect to a new scalar product (recall that $\hat{B}$ is bounded)
\be
  (\cdot\,|\,\cdot)_{\hat B}\ :=\ (\cdot\,|\,\hat{B}\cdot)_{\rm gr} \ .
\ee
The Hilbert space obtained  from ${\cal H}_{\rm gr}$ endowed with this new scalar product 
will be further denoted as ${\cal H}_{{\rm gr},\hat{B}}$.

In the discrete case one can retrieve the relation between scalar products almost 
immediately, namely, given $c$, one can show that the set of vectors 
$\{e_{\ppi_n(c),c}\in{\cal H}_{\rm gr}\,|\,n\in\mathbb{N}\}$ turns out to be orthogonal 
in ${\cal H}_{{\rm gr},\hat{B}}$, as
\be
  (e_{\ppi_n(c),c}|\hat{B}e_{\ppi_{n'}(c),c})_{\rm gr}\ 
  =\ \frac{1}{\ppi_n(c)}\frac{\rd c_n}{\rd\ppi}\Big{|}_{\ppi=\ppi_n(c)}\delta_{n,n'},
\ee
and the scalar product $(\cdot\,|\,\cdot)_c$ can be reexpressed in the following way,
\be\label{()cI}
  (e_{\ppi_n(c),c}|e_{\ppi_{n'}(c),c})_c \ 
  =\ (e_{\ppi_n(c),c}|\ppi_{n'}(c)\hat{B}e_{\ppi_{n'}(c),c})_{\rm gr} \ .
\ee

The continuous case is a bit more complicated. There, the analogy would be complete 
if it was true that $(e_{\ppi,c}|\hat{B}e_{\ppi',c})_{\rm gr}$ equal 
$\frac{1}{\ppi}\delta(\ppi-\ppi')$. It is not the case, though. Instead, the derivation 
of the product reveals the following result
\be
  (e_{\ppi,c}|\hat{B}e_{\ppi',c})_{\rm gr}\ 
  =\  2b\frac{\sin(a(\ppi,c)-a(\ppi',c))}{\ppi^2-\ppi'^2} \ .  \label{eBe'II} 
\ee
It follows then, that again each $e_{\ppi,c}\in {\cal H}_{{\rm gr},\hat{B}}$ and
the product equals
\be 
  (e_{\ppi,c}|\hat{B}e_{\ppi,c})_{\rm gr}\ 
  =\  \frac{b}{\ppi}\frac{\partial a(\ppi,c)}{\partial \ppi} \ .\label{eBeII}
\ee

Combining (\ref{eBe'II},\ref{eBeII}) with the definition (\ref{sprcII}) of the scalar 
product $(\cdot\,|\,\cdot)_c$ we find the desired relation
\be\label{()cII} 
  (e_{\ppi,c}|e_{\ppi',c})_{c} \ 
  =\ (e_{\ppi,c}|\ppi'\hat{B}e_{\ppi',c})_{\rm gr}\frac{1}{b}\,\delta(\sin(a(\ppi,c)-a(\ppi',c))) \ . 
\ee

The consequence of these results (\ref{()cI}) in the discrete and (\ref{()cII}) 
in the continuous case, is that the space of solutions to the quantum constraint 
$\hat{C}-c$ can be represented in terms of the unitary evolutions in the Hilbert space 
${\cal H}_{{\rm gr},\hat{B}}$. The continuous case however is essentially different 
than the discrete one. Let us now discuss the nature of the difference 
with a bit more detail.

In the discrete case, the Hilbert space ${\cal H}_c$ can be identified with 
(unitarily mapped onto) a subspace of the Hilbert space  ${\cal H}_{{\rm gr},\hat{B}}$, 
by means of  a unitary embedding
\be 
  e_{\ppi,c}\ \mapsto\ \sqrt{\ppi}\,e_{\ppi,c} \ .\label{e->ppie}
\ee
The operator $2\hat{B}^{-1}(\hat{C}_{\rm gr}-c\hat{\id})$ defined in the domain spanned by
the vectors $e_{\ppi_n(c),c}$, $n\in\mathbb{R}$, is essentially self adjoint, 
positive, and defines a unitary flow
\be 
  \mathbb{R}\ni T\mapsto U_c(T)\ :=\ \exp\left(iT \sqrt{2\hat{B}^{-1}(\hat{C}_{\rm gr}-c\hat{\id})}\right) \ . 
\ee
The space of solutions ${\cal H}^\pm_{{\rm kin},c}$ of positive/negative frequency 
becomes the space of functions
\be 
  T\ \mapsto\ U_c(\pm T)\psi(0) \ , \qquad 
  \psi(0)\in {\cal H}_c\subset {\cal H}_{{\rm gr},\hat{B}} \ .
\ee
The only technical subtlety is that we would usually expect ${\cal H}_{c}$ to be
the whole ${\cal H}_{{\rm gr},\hat{B}}$, whereas in our case there seems to be a 
possibility that ${\cal H}_{c}$ is a proper subspace.

In the continuous case,  $e_{\ppi,c}\in{\cal H}_{{\rm gr},\hat{B}}$ for every 
$\ppi>0$ and each $c\in\mathbb{R}$ despite of the fact, that it is normalizable 
to the Dirac delta with respect to $(\cdot\,|\,\cdot)_c$. Therefore, ${\cal H}_c$ 
can not be naturally identified with a subspace of  ${\cal H}_{{\rm gr},\hat{B}}$. 
It turns out however, that
the elements of ${\cal H}_c$ can be identified (via suitable unitary map) as families 
of vectors $(\psi^{(a)}\in{\cal H}_{c}^{(a)})_{a\in[0,\pi)}$ or, in other words, 
the formal integrals ${\int^{\oplus}}_{[0,\pi)} \rd a \psi^{(a)}$, where for every $a\in[0,\pi)$,
the Hilbert space  ${\cal H}_{c}^{(a)}$ is a suitable subspace of ${\cal H}_{{\rm gr},\hat{B}}$. 
Specifically, ${\cal H}_{c}^{(a')}$ is the completion of the subspace spanned by
\be 
  \{e_{\ppi,c}:\ a(\ppi,c)-a'=n\pi, n\in\mathbb{Z}\} \ .
\ee
The scalar product $(\cdot\,|\,\cdot)_c$ is equivalent to
\be 
  (\psi|\psi')\ =\ \int_{[0,\pi)} \rd a (\psi^{(a)}|\psi'^{(a)})_{\hat{B}} \ .
\ee
In order to write the exact unitary evolution map, we recall that the 
elements of ${\cal H}_c$ are formal integrals 
$\psi={\int^{\oplus}}_0^\infty \rd\ppi \psi(\ppi,c)e_{\ppi,c}$ with the scalar product
\be 
  (\psi|\psi')_c\ =\ \int_0^\infty \rd\ppi \overline{\psi_{\ppi,c}}\psi'_{\ppi,c} \ ,
\ee
thus for every $\psi={\int^\oplus}_0^\infty \rd\ppi\psi(\ppi,c)e_{\ppi,c}$ the family 
${\int^{\oplus}}_{[0,\pi)} \rd a\psi^{(a)}$ is given by
\be
  \psi^{(a')}\ =\ \sum_{\ppi:a(\ppi,c)-a'\in\pi \mathbb{Z}} 
  \left(\frac{\partial a(\ppi,c)}{\partial \ppi}\right)^{-\frac{1}{2}} \sqrt{b\ppi}\,\psi(\ppi,c) e_{\ppi,c}.
\ee

The operator $2\hat{B}^{-1}(\hat{C}_{\rm gr}-c\hat{\id})$, defined in this
case in the entire Hilbert space ${\cal H}_{{\rm gr},\hat{B}}$, in
each of the subspaces ${\cal H}_c^{(a)}\subset{\cal H}_{{\rm
gr},\hat{B}}$ becomes essentially self-adjoint and positive
definite. In consequence in each subspace ${\cal H}_c^{(a)}$ this operator
defines  a unitary evolution 
\be 
  U(T)_c^{(a)}\ =\ \exp \left(iT\sqrt{2\hat{B}^{-1}(\hat{C}_{\rm gr}-c\hat{\id})}\right) \ 
\ee
and positive/negative frequency solutions of the quantum constraint $\hat{C}-c$ 
are provided by a map 
\be 
  T\ \mapsto\ \int^\oplus_{[0,\pi)} \rd a\,\psi^{(a)}(T) \ ,\qquad 
  \psi^{(a)}(T)\ =\ U^{(a)}(\pm T)\psi^{(a)}(0)\in{\cal H}_c^{(a)} \ . 
\ee

To arrive to the characterization above the key idea is the consideration of the operator
$\hat{B}^{-1}(\frac{1}{2}\ppi^2\hat{B}+\hat{C}_{\rm gr}-c\hat{\id})$ instead of the original 
$\frac{1}{2}\ppi^2\hat{B}+\hat{C}_{\rm gr}-c\hat{\id}$. That lead us to the Hilbert space 
${\cal H}_{{\rm gr},\hat{B}}$. The unitary isometry
\be 
  \hat{B}^{\frac{1}{2}}\ :\ {\cal H}_{{\rm gr},\hat{B}}\ \rightarrow\ {\cal H}_{\rm gr}\label{B1/2}
\ee
can be used any time, to map all the Hilbert spaces ${\cal H}_c$, and, respectively, 
${\cal H}_c^{(a)}$ and the considerations therein, into subspaces $\tilde{\cal H}_c, \tilde{\cal H}_c^{(a)}\subset {\cal H}_{\rm gr}$, with the operator $2\hat{B}^{-1}(\hat{C}_{\rm gr}-c\hat{\id})$ carried  into the operator
$2\hat{B}^{-\frac{1}{2}}(\hat{C}_{\rm gr}-c\hat{\id})\hat{B}^{-\frac{1}{2}}$
and  the elements $e_{\ppi,c}\in{\cal H}_{{\rm gr},\hat{B}}$ mapped into $\tilde{e}_{\ppi,c}=\hat{B}^\frac{1}{2}e_{\ppi,c}\in{\cal H}_{\rm gr}$.

$\bullet$
final step of the construction is the derivation of the {\bf Dirac observables} 
in the Hilbert space ${\cal H}_{{\rm gr},c}$. Here, as the starting point, we use
eq. (\ref{qobsC}) which maps each operator $\hat{F}$ defined in 
${\cal H}_{\rm kin}=L^2(\mathbb{R})\otimes{\cal H}_{\rm gr}$ into  the Dirac observables 
$\hat{F}^{\rm D}_{(T,t)}$ depending on the value of the parameter $t$. The formula involves    
arbitrary symmetrization, defined up to the ordering ambiguity.

We fix the symmetrization, while analyzing Case I, requiring that it satisfies what follows.
\begin{itemize}
  \item[$i)$] The Dirac observable $f(\hat{\Pi})^{\rm D}_{(T,t)}$ corresponding to the operator 
    $f({\hat \Pi})\otimes \hat{\id}$ where $f$ is arbitrary function, is
    \be 
      f(\hat{\Pi})^{\rm D}_{(T,t)} \ =\ f({\hat \Pi}) \ .
    \ee  
  \item[$ii)$] For $\hat{F}\ =\ \hat{\id}\otimes \hat{G}$ the matrix element of the resulting observable 
    $\hat{G}^{\rm D}_{(T,t)}$ between states $\psi,\psi'\in {\cal H}^{\pm}_{{\rm kin},c}$ 
    represented (via the map \eqref{B1/2}) by
    \begin{subequations}\label{eq:matrix}\begin{align}
      T\ &\mapsto\ \tilde{\psi}(T)=U_c(\pm T)\tilde{\psi}(0)\in {\cal H}_{\rm gr} \ , &
      T\ &\mapsto\ \tilde{\psi}'(T)=U_c(\pm T)\tilde{\psi}'(0)\in {\cal H}_{\rm gr} \ ,
    \tag{\ref{eq:matrix}}\end{align}\end{subequations}
    equals
    \be  
      (\psi|\hat{G}^{\rm D}_{(T,t)}\psi')_{{\rm kin},c}\ 
      =\ (\tilde{\psi}(t)\,|\,\hat{G}\tilde{\psi}'(t))_{\rm gr} \ . 
    \ee
\end{itemize}
The last equality coincides here with the usual Schr\"odinger picture action of the operator 
${\hat G}$ at the instant $T=t$ on the  states $T\mapsto\tilde{\psi}(T)$ evolving in 
${\cal H}_{\rm gr}$. We consider this as an indication, that the chosen symmetrization 
is reasonable. The only subtlety is hidden in the fact, that the states are restricted 
to the subspace $\tilde{\cal H}_c$ of ${\cal H}_{\rm gr}$, hence the kinematical observable 
${\hat G}$ is in fact replaced by the projected observable $P\hat{G}P$, where $P$ is 
the orthogonal projection onto that subspace.

In the continuous case we apply the formula (\ref{qobsC}) analogously to the 
discrete one, additionally directly parachuting from it the fixing of the symmetrization 
ambiguities. It turns out that the relational observables mix the spaces ${\cal H}_c^{(a)}$ 
and ${\cal H}_c^{(a')}$ for every pair $a\not= a'$. For example, the observable 
$\hat{G}^{\rm D}_{(T,t)}$ corresponding to the kinematical observable 
$\hat{F}\ =\ \hat{\id}\otimes \hat{G}$ has the following matrix elements 
\be 
  ({\psi} \,|\, \hat{G}^{\rm D}_{(T,t)}{\psi}')_{{\rm kin},c}\ =\ \int_{[0,\pi)^2} \rd a
  \rd a'(\tilde{\psi}^{(a)}(t) \,|\, \hat{G}\tilde{\psi}^{(a')}(t))_{\rm gr}
\ee
between two states $\psi,\psi'\in {\cal H}^\pm_{{\rm kin},c}$ represented by
\be 
  T\mapsto\tilde{\psi}(T)\ =\ \int^\oplus_{[0,\pi)} \rd a \tilde{\psi}^{(a)}(T) \ ,\qquad
  T\mapsto \tilde{\psi}'(T)\ =\ \int^\oplus_{[0,\pi)} \rd a \tilde{\psi}'^{(a)}(T) \ .
\ee

In particular, even the identity observable $\hat{\id}\otimes \hat{\id}$ is mapped into 
$\hat{1}^{\rm D}_{(T,t)}$ such that
\be 
  ({\psi} \,|\, \hat{1}^{\rm D}_{(T,t)}{\psi}')_{{\rm kin},c}\ 
  =\ \int_{[0,\pi)^2} \rd a \rd a'(\tilde{\psi}^{(a)}(t)\tilde{\psi}^{(a')}(t))_{\rm gr} \ , 
\ee
which is inequivalent to $\hat{\id}\otimes \hat{\id}$. As we have explained in Section \ref{GAFRW}, 
this is an indication that in the corresponding classical theory, the reference function 
$T$ restricted to some physical trajectories does not achieve every value $t\in \mathbb{R}$, 
hence even the classical relational observable
\be 
  1^{\rm D}_{(T,t)}\ \not\equiv\ 1 \ .
\ee

Let us remind here, that in the construction outlined above we do not have to invoke the 
specific examples which exist in LQC. The only starting conditions are the assumptions 
listed at the beginning of this section.

The material presented above constitutes just a  sketch of the construction. 
The details of the derivation are provided in next two sections separately for the discrete
(Section \ref{Exp1}), and respectively, the continuous (Section \ref{Exp2}) case.

\section{The discrete case}\label{Exp1}

In this section we apply the framework  (\ref{Hkin}--\ref{psiqobspsi}) to the 
discrete case, which, up to technical details, has been defined in Section 
\ref{sec:res} (point \ref{it:discr-ass}). The detailed assumptions corresponding to this case are listed  
in the next subsection. As outlined in the previous section, our goal is the 
characterization of the solutions to the constraint, as well as the construction 
of the relational Dirac observables.

\subsection{Assumptions}\label{Exp1ass}

Suppose that the operators $\hat{C}_{\rm gr}, \hat{B}$ and $\hat{C}_\ppi$ \eqref{Cpi} satisfy 
the following assumptions:
\begin{enumerate}
  \item $0<\hat{B}< B_0\hat{\id}$, that is $\hat{B}$ is positive, bounded and invertible
    (the inverse may be unbounded)
  \item For every $\ppi\ge 0$, there is an orthonormal basis 
    $\{{e}_{\ppi,c_n(\ppi)}\in {\cal H}_{\rm gr}\}_{n\in \mathbb{N}}$, such that
    \be 
      \hat{C}_\ppi{e}_{\ppi,c_n(\ppi)}\ =\ c_n(\ppi) {e}_{\ppi,c_n(\ppi)} \ , 
    \ee
  \item $\ldots\ <\ c_n(0)\ <\ c_{n-1}(0)\ <\ \ldots\ <\ c_1(0)=:-c_0\ <0$ \label{it:ass-discr-bounds}.
  \item Each function  $\ppi\mapsto c_n(\ppi),$ is growing to infinity in the half 
    line $(0,\infty)$, and is differentiable.
  \item The functions $c_n(\ppi)$ corresponding to different $n$ never intersect,
    that is $\forall_{\ppi\geq 0}\ n\neq n'\ \Rightarrow\ c_n(\ppi)\neq c_n'(\ppi)$.
    \label{it:ass-discr-curves}
  \item The functions
    \be 
      \ppi\ \mapsto\ {e}_{\ppi,c_n(\ppi)}\in {\cal H}_{\rm gr} \ , \qquad
      c\ \mapsto\ e_{\ppi_n(c),c}\in {\cal H}_{\rm gr}
    \ee
    are continuous. \label{it:ass-discr-cont}
\end{enumerate}

\subsection{The spectral decomposition of the constraint operator $\hat{C}$.}
\label{Exm1decomp} 

Our starting point is the spectral decomposition of the operator $\hat{\Pi}$ 
provided on the abstract level in (\ref{Hpic}) followed by the (also abstract) 
decomposition of the operators $\hat{C}_\ppi$ \eqref{Cpi}. To start the first step 
(with respect to the description in Section \ref{sec:res}) in solving the quantum 
constraint, we derive from (\ref{Hpic}) a spectral decomposition (\ref{Hcpi}) of 
${\cal H}_{\rm kin}$ corresponding to the operator $\hat{C}$ (\ref{C}). In 
order to do so we apply the assumption of Section \ref{Exp1ass} and \eqref{psifurier} 
to explicitly define the measure in \eqref{Hpic}, which can be then expressed in 
the following, equivalent form
\be 
  \Psi(T)\ =\ \frac{1}{\sqrt{2\pi}}\int_{-\infty}^\infty
  \rd\ppi\sum_{n=1}^\infty {\psi}_{\ppi, c_n(\ppi)}e^{i\ppi T} {e}_{|\ppi|,c_n(\ppi)} \ ,
  \label{PiCdecomp1}
\ee
where ${\cal H}_{c,\ppi} = \mathbb{C}$. As explained in Section \ref{GAFRW}, it is 
still a decomposition of the operator $\hat{\Pi}$, further sub-decomposed 
with respect to the spectral decompositions of the operators $\hat{C}_\ppi$. 
However, the joint spectrum
\begin{equation}
  {\rm Spec}\ =\ \{(\ppi,c_n(\ppi))\,|\,\ppi\in\mathbb{R}, n\in \mathbb{N}\}
\end{equation} 
of a pair of operators is independent of the order of the decomposition. 
In this case the joint spectrum is the disjoint union  of curves
$\mathbb{R}\ni\ppi\mapsto (\ppi,c_n(\ppi))\in \mathbb{R}^2$, labelled by $n\in \mathbb{N}$. 
Therefore the Hilbert space ${\cal H}_{\rm kin}=L^2(\mathbb{R})\otimes{\cal H}_{\rm gr}$ 
is unitarily equivalent to $L^2({\rm Spec})$ with the measure $d\ppi \sum_n$, that is
\begin{equation}
  (\psi|\psi')_{\rm kin}\ 
  =\ \int_{-\infty}^\infty \rd\ppi \sum_n\overline{\psi_{\ppi,c_n(\ppi)}} \psi'_{\ppi,c_n(\ppi)} \ .
\end{equation}
To parametrize the $n$th curve (which accounts to {\rm Spec}) by the second eigenvalue: $c$, 
we split it into two branches, corresponding to $\ppi\ge 0$, and, respectively, $\ppi\le 0$,
\begin{equation}
  c\ \mapsto\ \begin{cases} 
              (\ppi_n(c),c) \ ,& \ppi\ge 0 \ ,\\
              (-\ppi_n(c),c) \ ,& \ppi\le 0 \ .
            \end{cases}\label{cparametr}
\end{equation}
where $\ppi_n(c)$ is the inverse function to $\ppi\mapsto c_n(\ppi)$
on the domain $\ppi\ge 0$. Using this change of variables we find
\be 
  (\psi|\psi')_{\rm kin}\ =\ \int_{-\infty}^\infty \rd c \sum_{n\ge n_c} 
  \left|\frac{\rd\ppi_{n}}{\rd c}\right| \left( \overline{\psi_{\ppi_n(c), c}}
  \psi'_{\ppi_n(c), c}\ +\ \overline{\psi_{-\ppi_n(c), c}}
  \psi'_{-\ppi_n(c), c}\right) \ ,\label{cprod}
\ee
where $n_c$ is the lowest value of $n\in\mathbb{N}$ such that $c\in c_n(\mathbb{R})$
(in the neighborhood of $c=0$ relevant for us, we have $n_c=1$).

Applying the same change of variables in \eqref{PiCdecomp1},
we arrive to the explicit spectral decomposition of the operator $C$ 
encoded in the formula
\be 
  \Psi(T)\ =\ \frac{1}{\sqrt{2\pi}}\int \rd c \sum_{n\ge n_c}
  \left|\frac{\rd\ppi_{n}}{\rd c}\right| \left( {\psi}_{\ppi_n(c) c}e^{i\ppi_n(c) T}\ +\
  {\psi}_{-\ppi_n(c) c}e^{-i\ppi_n(c) T}\right)
  {e}_{\ppi_n(c),c} \ .\label{cdec}
\ee
Note, that there are degenerate points in this decomposition at which
$\frac{\rd\ppi_n(c)}{\rd c}\rightarrow \infty$, which happens whenever
$\ppi_n(c)=0$. These points may seriously affect the well-definiteness of 
a function
\begin{equation}
  \left|\frac{\rd\ppi_{n}}{\rd c}\right| {\psi}_{\pm\ppi_n(c) c}e^{\pm i\ppi_n(c) T}
  {e}_{\ppi_n(c),c} \ ,
\end{equation}
at $\ppi_n(c)=0$. This is however not a problem in our case, as, due to the 
assumption \ref{it:ass-discr-bounds}, no $\ppi_n(c)$ vanishes in a sufficiently narrow 
neighbourhoud of $c=0$, the point in the spectrum we are interested in. 
For the sake of generality however, we will address this issue later on (see the 
remark at the end of Section \ref{Exp1prod}).

\subsection{Solutions to the constraint $\hat{C}-c\hat{\id}$.}\label{Exm1char}

At this point, having at our disposal (\ref{cprod},\ref{cdec}) we are in 
a position to start a characterization of the Hilbert space of solutions to the 
constraint defined by the constraint operator $\hat{C}$, that is the Hilbert space 
${\cal H}_{{\rm kin},0}$. Unfortunately, the measure $dc$ in (\ref{cdec}) is
that of Lebesgue, therefore the point $c=0$ of the spectrum of the
operator $\hat{C}$ is of measure $0$. Therefore, as mentioned in Section 
\ref{GAq}, a characterization of the corresponding Hilbert space 
${\cal H}_{{\rm kin},c=0}$ in the spectral decomposition will be meaningful only 
in the sense of a certain limit as $c\rightarrow 0$.

To specify it, we start by fixing arbitrary $c$ in the neighborhood $(-c_0,c_0)$ of $0$, 
and characterizing the Hilbert space ${\cal H}_{{\rm kin},c}$ of the solutions to a constraint 
operator $\hat{C}-c\hat{\id}$. This completes the $1$st step listed in Section \ref{sec:res}.
The Hilbert space ${\cal H}_{{\rm kin},c}$ defined by the decomposition
(\ref{cdec}) is the linear span of a set
$\{E_{\pm\ppi_n(c),c,n}\,|\,n\in\mathbb{N}\}$\footnote{The value $c_0$ is chosen to 
  be small enough to ensure that for all $c\in(-c_0,c_0)$ the lowest $\ppi_n(c)$ 
  \eqref{cparametr} always corresponds to the same curve specified in assumption 
  \ref{it:ass-discr-curves} of Section \ref{Exp1ass}.}
of ${\cal H}_{\rm gr}$ valued functions of the variable $T$ defined via \eqref{Tmapstoe}
\begin{equation}\label{psipm}
  E_{\pm\ppi_n(c),c}(T)\ =\ \frac{1}{\sqrt{2\pi}}e^{\pm i\ppi_n(c) T}e_{\ppi_n(c),c} \ ,
\end{equation}
and the scalar product $(\cdot|\cdot)_{{\rm kin},c}$ is such that
the above set of the functions is orthogonal, and satisfies
\begin{equation}
  (E_{\ppi_n(c),c}\,|\,E_{\ppi_{n}(c),c})_{{\rm kin},c}\ =\
  \left. \frac{dc_n}{d\ppi} \right|_{\ppi=\ppi_n(c)}\ 
  =\ (E_{-\ppi_n(c),c}\,|\,E_{-\ppi_{n'}(c),c})_{{\rm kin},c} \ .
\end{equation}
We can split the Hilbert space into positive and negative frequency sectors
\begin{equation} 
  {\cal H}_{{\rm kin},c}\ =\ {\cal H}^+_{{\rm kin},c}\oplus 
  {\cal H}^-_{{\rm kin},c} \ ,  \label{Hpm}
\end{equation}
where ${\cal H}^+_{{\rm kin},c}$ (${\cal H}^-_{{\rm kin},c}$) is
spanned by the functions $E_{\ppi_n(c),c}$ ($E_{-\ppi_n(c),c}$). The
scalar product in each of these sectors  
can be expressed by a suitably modified scalar product in the Hilbert space 
${\cal H}_{\rm gr}$. Indeed, let us consider the vector subspace
\be
  {\rm Span}(\,e_{\ppi_n(c),c}\,|\,n\in \mathbb{N}\,)\subset {\cal H}_{\rm gr} 
\ee
and endow it with a new scalar product $(\cdot|\cdot)_c$, such that
\be 
  (e_{\ppi_n(c),c} |e_{\ppi_{n'}(c),c} )_{c}\ 
  =\ 2\pi\left|\frac{\rd\ppi_{n}}{\rd c}(c)\right|^{-1}\delta_{nn'} \ .\label{cprod2}
\ee
Also, denote
\be 
  {\cal H}_c\ =\ \overline{{\rm Span}(\,e_{\ppi_n(c),c}\,|\,n\in\mathbb{N}\,)} \ . 
\ee
Then, for every $\psi^\pm_1,\psi_2^\pm\in {\cal H}^\pm_{{\rm kin},c}$, the scalar 
product $(\psi^\pm_1\,|\,\psi_2^\pm)_{{\rm kin},c}$ can be expressed by the 
${\cal H}_c$ values $\psi^\pm_1(T),\psi_2^\pm(T)$ at instant $T$ in the following way
\be 
  (\psi^\pm_1\,|\,\psi_2^\pm)_{{\rm kin},c}\ 
  =\ (\psi^\pm_1(T)\,|\,\psi_2^\pm(T))_{c} \ , 
\ee
where the right hand side is independent of the choice of the value
of the $T$ variable.

Due to assumption \ref{it:ass-discr-cont} of Section \ref{Exp1ass} all the 
elements of the construction are continuous with respect to $c$. The Hilbert 
space of solutions to the quantum constraint defined by the operator $\hat{C}$ is 
then given just by setting $c=0$.

\subsection{The scalar product between the solutions} \label{Exp1prod}

The scalar product $(\cdot|\cdot)_{c}$ has been introduced in the subspace 
${\rm Span}(e_{\ppi_n(c),c}\,|\,n\in\mathbb{N})\subset {\cal H}_{\rm gr}$
by declaring its matrix in the basis $\{e_{\ppi_n(c),c}\,|\,n\in\mathbb{N}\}$. 
It appears however, (as we show below) that this product can be also 
defined in a compact way, namely,
\be 
  (\cdot|\cdot)_{c}\ =\ 2\pi(\cdot|\hat{B}\hat{\Pi}_c\cdot)_{\rm gr} \ , \label{scalprodc}
\ee
where $\hat{\Pi}_c$ is an operator defined in ${\rm Span}(e_{\ppi_n(c),c}\,|\, n\in\mathbb{N})$ 
by
\be 
  \hat{\Pi}_c e_{\ppi_n(c),c}\ :=\ \ppi_n(c)e_{\ppi_n(c),c} \ .\label{Pic}
\ee
Let us now derive the relation \eqref{scalprodc}, thus realizing the $3$rd step
outlined in Section \ref{sec:res}. To start with, let us substitute into the identity
\be 
  (\hat{C}_{\rm gr}e_{\ppi,c_n(\ppi)}\,|\,e_{\ppi',c_{n'}(\ppi')})\ 
  =\ (e_{\ppi,c_n(\ppi)}\,|\,\hat{C}_{\rm gr}e_{\ppi',c_{n'}(\ppi')}) 
\ee 
valid for $\ppi,\ppi'\ge 0$, the condition (which also shows, that each 
$e_{\ppi,c_n(\ppi)}\in{\cal H}_{\rm gr}$ is in the domain (after a closure of the operator)
of $\hat{C}_{\rm gr}$ as $\hat{B}$ is boundend)
\be 
  \hat{C}_{\rm gr}e_{\ppi,c_n(\ppi)}\ =\ (c_n(\ppi)\hat{\id} -
  \frac{1}{2}\ppi^2\hat{B})e_{\ppi,c_n(\ppi)} \ . 
\ee
The result of this operation is
\begin{equation}
  (c_n(\ppi) - c_{n'}(\ppi')) ({e}_{\ppi,c_n(\ppi)}\,
  |{e}_{\ppi',c_{n'}(\ppi')})\ =\
  \frac{1}{2}(\ppi^2-\ppi'^2)({e}_{\ppi,c_n(\ppi)} |\hat{B}
  {e}_{\ppi',c_{n'}(\ppi')}) \ .
\end{equation}
which implies in particular, that
\begin{equation}
  ({e}_{\ppi_n(c),c}|\hat{B}{e}_{\ppi_{n'}(c),c})\ =\ 0 \ , \quad 
  {\rm whenever} \quad n\not=n' \ .\label{scalprod0}
\end{equation}
Furthermore,
\be
  ({e}_{\ppi,c_n(\ppi)} |\hat{B} {e}_{\ppi,c_{n}(\ppi)})\ 
  =\ \lim_{\ppi'\rightarrow\ppi}2\frac{c_n(\ppi) 
  - c_{n}(\ppi')}{\ppi^2-\ppi'^2} ({e}_{\ppi,c_n(\ppi)} |
  {e}_{\ppi',c_{n}(\ppi')})
  =\ \frac{1}{\ppi}\frac{\rd c_n}{\rd\ppi}
\ee
and in terms of the parametrization by (\ref{cparametr}), the product equals
\be 
  \ppi_n(c)({e}_{\ppi_n(c),c} |\hat{B} {e}_{\ppi_n(c),c})\ 
  =\ (\frac{\rd\ppi_n(c)}{\rd c})^{-1}\label{scaprod2} \ .
\ee
The comparison with the scalar product $(\cdot|\cdot)_{c}$ of (\ref{cprod2}) 
gives then (\ref{scalprodc}) as we stated at the beginning of this subsection.

The operator $\hat{\Pi}_c$ can be expressed by the operators $\hat{B}$ and 
$\hat{C}_{\rm gr}$. We have
\be 
  2\hat{B}^{-1}(c\hat{\id}-\hat{C}_{\rm gr})e_{\ppi_n(c),c}\ =\
  \ppi_n^2(c)e_{\ppi_n(c),c}\ =\ \Pi_c^2e_{\ppi_n(c),c} \ .
\ee
Therefore in the Hilbert ${\cal H}_c$  defined by completing 
${\rm Span}(e_{\ppi_n(c),c}\,|\, n\in\mathbb{N})\subset{\cal H}_{\rm gr}$ with
respect to $(\cdot|\cdot)_c$ in which the operator $2\hat{B}^{-1}(c\hat{\id}-\hat{C}_{\rm gr})$ 
is self-adjoint, we can write
\be 
  \hat{\Pi}_c\ =\ \sqrt{2\hat{B}^{-1}(c\hat{\id}-\hat{C}_{\rm gr})} \ ,\label{Pic'}
\ee
and conclude that the scalar product $(\cdot|\cdot)_c$ is
\be 
  (\cdot|\cdot)_c\ =\ 2\pi(\cdot|\hat{B}\sqrt{2\hat{B}^{-1}(c\hat{\id}-\hat{C}_{\rm gr})}\cdot)_{\rm gr} \ .
  \label{prodcfinal} 
\ee
To make the relation with (\ref{3a}) closer, we can endow the vector subspace
${\rm Span}(e_{\ppi_n(c),c}\,|\, n\in\mathbb{N})$ of ${\cal H}_{\rm gr}$ with
a scalar product $(\cdot|\hat{B}\cdot)$ and in that auxiliary Hilbert space
the operator    $2\hat{B}^{-1}(c\hat{\id}-C_{\rm gr})$ is also self adjoint and positive,
so we can understand its  square root in (\ref{prodcfinal}) in the sense of this scalar product.\\

\noindent{{\bf Remark:}} The spectral decomposition formula (\ref{cdec}) has degenerate 
points such that
\be\label{eq:diff-deg}
  \frac{\rd c_n}{\rd\ppi}\ =\ 0 \ .
\ee
which are present at
\be
  \ppi\ =\ 0 \ , \quad c\ =\ c_n(0) \ , \quad n\in\mathbb{N} \ .
\ee
The corresponding functions $E_{0,c_n(0)}$, $n\in\mathbb{N}$ may be called ``the zero modes'', 
because of the vanishing frequency $\ppi$. The difficulty related to the vanishing of the 
lefthand side of \eqref{eq:diff-deg} is not relevant for us,as due to the assumption that 
$c_n(0)\not= 0$, there is no zero mode among the solutions to the quantum constraint 
$\hat{C}-c\hat{\id}$ (provided $c$ is sufficiently close to $0$). Nonetheless, 
let us consider in this Remark, the quantum constraint operator is $\hat{C}-c_n(0)\hat{\id}$, 
to see if there is a natural extension/limit of our framework, as $c\rightarrow c_n(0)$. 
This happens to be indeed the case, as each zero mode $E_{0,c_n(0)}$ is the right 
limit (in the sense of the $L^\infty$  topology in the space of functions
$\mathbb{R}\rightarrow {\cal H}_{\rm gr}$)
\be
  E_{0,c_n(0)}\ =\ \lim_{c\searrow c_n(0)}E_{\pm \ppi_n(c),c} \ ,
\ee
taken along the curves considered in assumption \ref{it:ass-discr-curves}. In consequence
a natural definition of its norm is
\be \label{eq:deg-prod}
  (E_{0,c_n(c)}|E_{0,c_n(0)})_{{\rm kin},c_n(0)}\ 
  :=\ \lim_{c\searrow c_n(0)}(E_{\pm \ppi_n(c),c}| E_{\pm \ppi_n(c),c})_{{\rm kin},c}\ 
  =\ \frac{\rd c_n}{\rd\ppi}(0)\ =\ 0 \ ,
\ee
that solves the problem of the zero modes. 
One should however remember, that potentially there may exist inequivalent ways
of taking the zero mode limit, giving in principle the result different than 
\eqref{eq:deg-prod}.

\subsection{The Dirac observables}\label{diracc1}

As discussed already in the previous part of the article, the operator 
$\hat{\Pi}\otimes \hat{\id}$ defined in
${\cal H}_{\rm kin}=L^2(\mathbb{R}\otimes{\cal H}_{\rm kin})$ is a
quantum Dirac observable, so is any operator of the form $f(\hat{\Pi})\otimes \hat{\id}$.

Another class of the Dirac observables can be constructed by the
relational observable  method (see (\ref{qobsC})) from any operator of the form
\be 
  \hat{F}\ =\ \hat{\id}\otimes\hat{G}
\ee
where $\hat{G}$ is an operator in ${\cal H}_{\rm gr}$.

The starting point of the construction is the integral (\ref{qobsC}), that is
\be 
  \hat{F}^{D}_{(T,t)}\ =\ \int d\tau e^{-i\tau \hat{C}}\circ {\rm Sym} \left(\,|\,\hat{\Pi}\otimes \hat{B}
  \,|\,\hat{F}\circ \delta(\hat{T}-t\hat{\id})\otimes \hat{\id}\right)\circ e^{i\tau\hat{C}} \ .
\ee 
This formula is defined up to an ambiguity in the symmetrization ``Sym'', for which
we propose some natural choice, which we introduce in two steps 
\begin{equation} 
  {\rm Sym}\ =\ {\rm Sym}_2\circ {\rm Sym}_1 
\end{equation}
as the composition of two operations:
\begin{enumerate}[a)]
  \item the first one is a symmetrization with respect to $\hat{\Pi}\otimes \hat{B}$
    \be
      {\rm Sym}_1\left(\,|\,\hat{\Pi}\otimes \hat{B}
      \,|\,\circ\hat{A}\right)\ =\ \sqrt{\,|\,\hat{\Pi}\otimes \hat{B}
      \,|\,}\circ \hat{A}\circ \sqrt{\,|\,\hat{\Pi}\otimes \hat{B}
      \,|\,} \ ,
    \ee
  \item the second one is
    \begin{equation} 
      {\rm Sym}_2 \hat{A}\ =\ \theta(\hat{\Pi})\hat{A} \theta(\hat{\Pi}) + \theta(-\hat{\Pi})\hat{A}
      \theta(-\hat{\Pi}) \ ,
    \end{equation}
    where $\theta$ is a Heaviside step function
    \be 
      \theta(\ppi)\ =\ \begin{cases}
                         1 \ , &  \text{if } \ppi>0 \ , \\
                         0 \ , & \text{otherwise}.
                       \end{cases}
    \ee
    Its classical counterpart is an identity for every $F$ such that
    $ F(T,\Pi=0)\ =\ 0$.
\end{enumerate}
The resulting combined symmetrization is of the form
\be
  {\rm Sym }\left(\,|\,\hat{\Pi}\otimes \hat{B}
  \,|\,\circ\hat{A}\right)\ =\theta(\hat{\Pi})\sqrt{\,|\,\hat{\Pi}\otimes \hat{B}
  \,|\,}\hat{A}\sqrt{\,|\,\hat{\Pi}\otimes \hat{B}
  \,|\,}\theta(\hat{\Pi})+\theta(-\hat{\Pi})\sqrt{\,|\,\hat{\Pi}\otimes \hat{B}
  \,|\,}\hat{A}\sqrt{\,|\,\hat{\Pi}\otimes \hat{B}
  \,|\,}
  \theta(-\hat{\Pi}) \ .
\ee

The component ${\rm Sym}_{2}$ is fixed via imposing a simple consistency 
condition, while considering a simple example
\be 
  \hat{F}\ =\ f(\hat{\Pi})\otimes \hat{\id} \ ,
\ee
and the corresponding relational observable ${f(\hat{\Pi})}^{\rm D}_{(T,t)}$.
In the classical theory, given an observable $\Gamma\ni\gamma=(T,\Pi,\ldots)\mapsto f(\Pi)$, 
the corresponding relational  observable ${f}^D_{T,t}$ (see (\ref{classobs})) 
equals $f$ except for $\Pi=0$.
In the case at hand, however, ${\cal H}_{\rm kin,\hat{\Pi}\not= 0}={\cal H}_{\rm kin}$ 
therefore the condition reads\footnote{In the case at hand, $\ppi_n(c=0)\not= 0$ 
  therefore the part $\ppi=0$ of the spectrum is irrelevant.}
\be 
  f(\hat{\Pi})^{\rm D}_{(T,t)}\ =\ f(\hat{\Pi}) \ . 
\ee
To arrive to this result, the choice of the splitting 
$1=\theta(\Pi)+\theta(-\Pi)$ is crucial. Otherwise, there would be mixing between 
negative and positive eigenvalues $\ppi$. Indeed, with this choice, due to the formula
(\ref{scalprod0},\ref{scaprod2})
\begin{subequations}\label{GDsesqui}\begin{align} 
  (\psi\,|\,{f(\hat{\Pi})}^D_{(T,t)}\psi')_{\rm kin}\ 
  &=\ \int \rd c \sum_{n,n'\ge n_c}\frac{\rd\ppi_n(c)}{\rd c}\frac{\rd\ppi_{n'}(c)}{\rd c}
  |\ppi_n(c)\ppi_{n'}(c)|^{\frac{1}{2}}\left( e_{\ppi_n(c),c}\,| \hat{B}\, 
    e_{\ppi_{n'}(c),c}\right)_{\rm gr}\cdot \tag{\ref{GDsesqui}} \\
  &\cdot(f(\ppi_{n'}(c)) e^{-it(\ppi_n(c)-\ppi_{n'}(c))}\overline{\psi_{\ppi_n(c),c}}\psi'_{\ppi_{n'}(c),c}\
  +\ f(-\ppi_{n'}(c))e^{it(\ppi_n(c)-\ppi_{n'}(c))}\overline{\psi_{-\ppi_n(c),c}}\psi'_{-\ppi_{n'}(c),c}) 
  \notag \\ 
  \ &=\  \int dc \sum_{n\ge n_c} \left(f(\ppi_n(c))\overline{\psi_{\ppi_n(c),c}}\psi'_{\ppi_{n}(c),c}\ 
  +\ f(-\ppi_n(c))\overline{\psi_{-\ppi_n(c),c}}\psi'_{-\ppi_{n}(c),c}\right)\ 
  =\  (\psi|f(\hat{\Pi})\psi')_{\rm kin} \ , \notag
\end{align}\end{subequations}
as required. In particular, if $f$ is identically $1$, we have
\be 
  \hat{1}^{\rm D}_{(T,t)}\ =\ \hat{\id} \ .
\ee

The first component ${\rm Sym}_1$ of the symmetrization can be tested on a kinematical 
observable of the form
\be
  \hat{F}\ =\ \hat{\id}\otimes\hat{G} \ .
\ee
The relational Dirac observable $\hat{G}^{D}_{(T,t)}$ corresponding to it takes the form
\begin{equation}\begin{split}
  \hat{G}^{D}_{(T,t)}\ =\ \int \rd\tau e^{-i\tau \hat{C}}&\circ \big( \theta(\hat{\Pi})|\hat{\Pi}|^{\frac{1}{2}}
  \delta(\hat{T}-t\hat{\id})|\hat{\Pi}|^{\frac{1}{2}} \theta(\hat{\Pi})\  \\
  &\hphantom{+}\quad +\ \theta(-\hat{\Pi})|\hat{\Pi}|^{\frac{1}{2}}
  \delta(\hat{T}-t\hat{\id})|\hat{\Pi}|^{\frac{1}{2}} 
  \theta(-\hat{\Pi}) \big)\otimes \sqrt{\hat{B}}\,\hat{G}\sqrt{\hat{B}}\circ
  e^{i\tau\hat{C}} \ . \label{GDTt}
\end{split}\end{equation}
We will see in the next subsection, that this choice is distinguished by a representation 
of each Hilbert space ${\cal H}^\pm_{{\rm kin},c}$  as the space of functions 
$T\mapsto \tilde{\psi}(T)\in{\cal H}_{\rm gr}$.

In terms of the spectral decomposition (\ref{cprod},\ref{cdec}),
$\hat{G}^D_{(T,t)}$ defines  in ${\cal H}_{\rm kin}$ a sesquilinear form
\begin{equation}\begin{split} 
  (\psi\,|\,\hat{G}^D_{(T,t)}\psi')_{\rm kin}\ 
  =\ &\int \rd c \sum_{n,n'\ge n_c}\frac{\rd\ppi_n(c)}{\rd c}\frac{\rd\ppi_{n'}(c)}{\rd c}
  |\ppi_n(c)\ppi_{n'}(c)|^{\frac{1}{2}}\left( e_{\ppi_n(c),c}\,|
  \hat{B}^{\frac{1}{2}}\hat{G}\hat{B}^{\frac{1}{2}}\,
  e_{\ppi_{n'}(c),c}\right)_{\rm gr}\cdot \\
  &\cdot(e^{-it(\ppi_n(c)-\ppi_{n'}(c))}\overline{\psi_{\ppi_n(c),c}}\psi'_{\ppi_{n'}(c),c}\
  +\ e^{it(\ppi_n(c)-\ppi_{n'}(c))}\overline{\psi_{-\ppi_n(c),c}}\psi'_{-\ppi_{n'}(c),c})
  \label{GDsesqui-med}
\end{split}\end{equation}
defined by an operator in ${\cal H}_{\rm kin}$ which in terms of the spectral decomposition 
(\ref{cdec}) ${\cal H}_{\rm kin}=\int^\oplus{\cal H}_{{\rm kin},c}$ can be expressed as 
a family of the operators $(\hat{G}^D_{(T,t)c})_{c\in\mathbb{R}}$. Indeed, given two 
$\psi,\psi'\in  {\cal H}_{{\rm kin},c}$,
\begin{subequations}\begin{align}
  \psi\ &=\ \sum_{n\ge n_n}(\psi^+_n e^{i\ppi_n(c)T}e_{\ppi_n(c),c}\ +\ \psi^-_n e^{-i\ppi_n(c)T}e_{\ppi_n(c),c})  \ , \\
  \psi'\ &=\ \sum_{n\ge n_n}(\psi'^+_n e^{i\ppi_n(c)T}e_{\ppi_n(c),c}\ +\ \psi'^-_n e^{-i\ppi_n(c)T}e_{\ppi_n(c),c}) \ ,
\end{align}\end{subequations}
we have
\begin{equation}\begin{split}
  (\psi\,|\,\hat{G}^D_{(T,t)c}\psi)_{{\rm kin},c}\ 
  =\ \sum_{n,n'\ge n_c}&\frac{\rd\ppi_n(c)}{\rd c}\frac{\rd\ppi_{n'}(c)}{\rd c}
  |\ppi_n(c)\ppi_{n'}(c)|^{\frac{1}{2}}\left( e_{\ppi_n(c),c}\,|
  \hat{B}^{\frac{1}{2}}\hat{G}\hat{B}^{\frac{1}{2}}\,
  e_{\ppi_{n'}(c),c}\right)_{\rm gr}\cdot\\
  &\cdot(e^{-it(\ppi_n(c)-\ppi_{n'}(c))}\overline{\psi^+_{n}}\psi'^+_{n'}\
  +\ e^{it(\ppi_n(c)-\ppi_{n'}(c))}\overline{\psi^-_{n}}\psi'^-_{n'}) \ .
\end{split}\label{GDsesqui-dec}\end{equation}
This implies in particular, that the operator preserves the positive/negative 
frequency subspaces ${\cal H}^\pm_{{\rm kin},c}$.

The above formula defining the observable $\hat{G}^D_{(T,t)c}$ appears quite complicated.
In the next subsection we will see, that this operator can be expressed in much more compact 
form, once we view the solutions to the constraint as unitarily evolving states in ${\cal H}_{\rm gr}$.

\subsection{The constraint as evolution  in ${\cal H}_{\rm gr}$}\label{Exp1evolution} 

The analysis presented in Section \ref{Exm1char} has shown, that the solutions to the
quantum constraint defined by the operator $\hat{C}-c\hat{\id}$ form the Hilbert space 
${\cal H}^-_{{\rm kin},c}\oplus{\cal H}^+_{{\rm kin},c}$ (\ref{Hpm}), and the subspace
${\cal H}^\pm_{{\rm kin},c}$ consists of functions given by (\ref{psipm}).
Furthermore, we have found that ${\cal H}^\pm_{{\rm kin},c}$  can be characterized as
the Hilbert space of functions
\be 
  T\mapsto \psi(T)\ \in\ {\cal H}_c\ =\ \overline{{\rm Span}(e_{\ppi_n(c),c}\,|\,n\ge n_c)}\label{Hc}
\ee
endowed with the scalar product $(\cdot|\hat{B}\sqrt{2\hat{B}^{-1}(c\hat{\id}-\hat{C}_{\rm gr})}\cdot )_{\rm gr}$.
These functions are solutions to the equation
\be 
  \frac{1}{i}\frac{\rd\psi^{\pm}(T)}{\rd T}\ =\ \pm
  \sqrt{2\hat{B}^{-1}(c\hat{\id}-\hat{C}_{\rm gr})}\,\psi^{\pm}(T) \ ,\label{schroed} 
\ee
and the scalar product between two of them (denoted here as $\psi^\pm$ and $\psi'^\pm$) is
\be\
  (\psi^\pm\,|\,\psi'^\pm)_{{\rm kin},c}\ 
  =\  (\psi^\pm(T)|\hat{B}\sqrt{2\hat{B}^{-1}(c\hat{\id}-\hat{C}_{\rm gr})}\,\psi'^\pm(T))_{\rm gr} \ ,
\ee
where the righthand side is evaluated at any instant of $T$.

In consequence, the Hilbert space $\Hil_c$ is constructed
from the subspace of ${\cal H}_{\rm gr}$ endowed with the new
scalar product \eqref{scalprodc}. In the Hilbert space ${\cal H}_c$ the
operator $2\hat{B}^{-1}(c\hat{\id}-\hat{C}_{\rm gr})$ is diagonal in the orthogonal basis
$\{e_{\ppi_n(c),c}\, ,\, n\ge n_c\}$ and thus self-adjoint. More precisely, 
a solution to (\ref{Hc}) is defined by any ``initial data'' $\psi(0)\in {\cal H}_c$ 
and the formula
\be 
  \psi^\pm(T)\ =\ {\rm exp}(\pm i T\sqrt{2\hat{B}^{-1}(c\hat{\id}-\hat{C}_{\rm gr})})\,\psi(0) \ .
  \label{sol1}
\ee

To view that evolution as defined directly in the Hilbert space ${\cal H}_{\rm gr}$,  
we use a unitary embedding
\be 
  \hat{B}^{\frac{1}{2}}(2\hat{B}^{-1}(c\hat{\id}-\hat{C}_{\rm gr}))^{\frac{1}{4}}\ 
  :\ {\cal H}_c\ \rightarrow {\cal H}_{\rm gr} \ ,
\label{unitary}\ee
image of which we denote by $\tilde{\cal H}_c\subset {\cal H}_{\rm gr}.$

That embedding maps the solutions to (\ref{schroed}) into solutions
$\tilde{\psi}^\pm(\cdot)$ to the transformed equation
\be 
  \frac{1}{i}\frac{\rd\tilde{\psi}^{\pm}(T)}{\rd T}\ 
  =\ \pm \sqrt{2\hat{B}^{-\frac{1}{2}}(c\hat{\id}-\hat{C}_{\rm gr})\hat{B}^{-\frac{1}{2}}}\,\tilde{\psi}^{\pm}(T)
  \label{schroed2-pre} \ . 
\ee
The domain of the operator $2\hat{B}^{-\frac{1}{2}}(c\hat{\id}-\hat{C}_{\rm gr})\hat{B}^{-\frac{1}{2}}$ 
in $\tilde{\cal H}_c$ is
\be 
  {\rm Span}(\hat{B}^{\frac{1}{2}}e_{\ppi_n(c),c}\,|\,n\ge n_c) \ . 
\ee
Via this map, each Hilbert space ${\cal H}^\pm_{{\rm kin},c}$ of
solutions to the constraint $(\hat{C}-c\hat{\id})$ is represented by functions
\begin{equation}\begin{split}
  \tilde{\psi}^\pm:\mathbb{R}\ 
  &\rightarrow\ \tilde{\cal H}_c\subset{\cal H}_{\rm gr} \ , \\
  \tilde{\psi}^\pm(T)\ 
  &=\ {\rm exp}(\pm i T\sqrt{2\hat{B}^{-\frac{1}{2}}(c\hat{\id}-\hat{C}_{\rm gr})\hat{B}^{-\frac{1}{2}}})\tilde{\psi}(0) \ ,
  \label{sol2}
\end{split}\end{equation}	
and the scalar product between them (evaluated at any instant of $T$) equals
\begin{equation}
  (\tilde{\psi}^\pm\,|\,\tilde{\psi}'\pm)\ =\ (\tilde{\psi}^\pm(T)\,|\,\tilde{\psi}'(T)\pm)_{\rm gr} \ .
\end{equation}

One has to note however, that we do not know whether for given $c$, the functions 
$\hat{B}^{\frac{1}{2}}e_{\ppi_n(c),c}$ span a dense subspace of ${\cal H}_{\rm gr}$. In consequence  
$\tilde{\cal H}_c$ may a priori be a proper subset of ${\cal H}_{\rm gr}$.
Nevertheless, the operator $2\hat{B}^{-\frac{1}{2}}(c\hat{\id}-\hat{C}_{\rm gr})\hat{B}^{-\frac{1}{2}}$ 
is still essentially self-adjoint as long, as we consider it in $\tilde{\cal H}_c$.

The relational Dirac observables (\ref{GDsesqui-med}) of the previous subsection have in this 
representation a compact form. To write it, we will use the abbreviation $\hat{\Pi}_c$ (\ref{Pic})
for the operator $\sqrt{2\hat{B}^{-1}(c\hat{\id}-\hat{C}_{\rm gr})}$, and
\be 
  \tilde{\Pi}_c\ :=\  \sqrt{2\hat{B}^{-\frac{1}{2}}(c\hat{\id}-\hat{C}_{\rm gr})\hat{B}^{-\frac{1}{2}}} \ .
\ee
The sesquilinear form (\ref{GDsesqui-dec}) is defined by an operator which preserves the spectral 
decomposition ${\cal H}_{\rm kin}=\int^\oplus \rd c\,{\cal H}^+_{{\rm kin},c}\oplus {\cal H}^-_{{\rm kin},c}$ of 
the constraint operator $\hat{C}$. The operator is given by a family of operators 
$(\hat{G}^{{\rm D}\pm}_{(T,t)c})_{c\in\mathbb{R}}$, $c\in\mathbb{R}$, defined, respectively, 
in ${\cal H}^\pm_{{\rm kin},c}$. Given two elements $\psi^\pm,\psi'^\pm\in {\cal H}^\pm_{{\rm kin},c}$,
represented by (\ref{sol1}) the sesquilinear form (\ref{GDsesqui-dec}) assigns the following number
\be 
  (\psi^\pm\,|\,\hat{G}^{{\rm D}\pm}_{(T,t)c}\psi'^\pm)_{{\rm kin},c}\ 
  =\ (\psi^\pm(t)\,|\,\hat{\Pi}_c^{-\frac{1}{2}}\hat{B}^{-\frac{1}{2}}_{\vphantom{c}}\hat{G}
  \hat{B}^{\frac{1}{2}}_{\vphantom{c}}\hat{\Pi}_c^\frac{1}{2}\psi^{\pm}(t))_{\rm gr} \ . 
\ee
Using the unitary embedding (\ref{unitary}) and denoting its action on $\psi^\pm,\psi'^\pm$ by $\tilde{\psi}^\pm,\tilde{\psi}'^\pm$ we can write this formula in even simpler form
\be 
  (\psi^\pm\,|\,\hat{G}^{{\rm D}\pm}_{(T,t)c}\psi'^\pm)_{{\rm kin},c}\ 
  =\ (\tilde{\psi}^\pm(t)\,|\,\hat{G}\tilde{\psi}^{\pm}(t))_{\rm gr} \ . 
\ee
Briefly speaking  the action of this operator consists in acting with $\hat{G}$ at the 
value of $\tilde{\psi}^\pm$ at $T=t$ and evolving the result. In this picture 
there is however a technical subtlety: the presence of the nontrivial orthogonal 
projection operator $\hat{P}:{\cal H}_{\rm gr}\ \rightarrow\ {\cal H}_c$. The precise form of the 
quantum relational observable operator in the tilded representation is
\begin{equation}
  (\hat{G}^{{\rm D}\pm}_{(T,t)c}\tilde{\psi}^\pm)(T)\ 
  =\ e^{\pm i (T-t)\tilde{\Pi}_c}\hat{P}\hat{G}\tilde{\psi}(t) \ .
\end{equation}

\subsection{Discussion, the limit $c\rightarrow 0$}

In previous subsections we have characterized the Hilbert space ${\cal H}_{{\rm kin},c}$ 
(corresponding to the fixed\footnote{In a suitable neighborhood of $0$.} value of $c$) in the 
spectral decomposition of the operator $\hat{C}$ and introduced the relational quantum Dirac 
observables therein.
The functions $E_{\ppi_n(c),c}$, $n\in\mathbb{N}$, (\ref{psipm}) are defined up to a rescaling
\be 
  E_{\pm\ppi_n(c),c}\ \mapsto\ e^{i\alpha(\pm\ppi_n(c),c)}E_{\pm\ppi_n(c),c} \ .
\ee
However, the definition of the Hilbert space ${\cal H}_{{\rm kin},c}$ and the  structures we 
introduced to characterize the resulting theory are invariant with respect to that transformation. 
Also, due to the continuity of the map
\be 
  (-c_0,\infty)\ \ni\ c\ \mapsto\ E_{\pm\ppi_n(c),c} \ ,
\ee
there is a notion of the continuity in $c$ of all the relevant structures, namely
of
\begin{enumerate}[(i)]
  \item ${\cal H}_{{\rm kin},c}$,
  \item $\tilde{\cal H}_c\subset {\cal H}_{\rm gr}$,
  \item $\tilde{\Pi}_c\ =\ \sqrt{\hat{B}^{-\frac{1}{2}}(c\hat{\id}-\hat{C}_{\rm gr}) \hat{B}^{-\frac{1}{2}}}$.
\end{enumerate}
Therefore, eventhough the point $c=0$ is of the measure zero, the continuity makes the Hilbert 
space ${\cal H}_{{\rm kin},c=0}$, its characterization and the relational quantum observables 
uniquely defined.

The result takes the appearance of two copies of the Schr\"odinger-like quantum mechanics,
whose states are pairs of elements $\tilde{\psi}^+,\tilde{\psi}^-\in \tilde{\cal H}_0\subset{\cal H}_{\rm gr}$, 
and their evolution (independent of each other) is governed by two the generalized Hamiltonian operators
$\pm\sqrt{-\hat{B}^{-\frac{1}{2}}\hat{C}_{\rm gr}\hat{B}^{-\frac{1}{2}}}$. The scalar field 
$T$ plays the role of time, and the relational Dirac observables derived from the kinematical 
observables, operators in ${\cal H}_{\rm gr}$ are just the operators pulled back from 
${\cal H}_{\rm gr}$ into the subspace ${\cal H}_{0}$ and acting on the solutions to 
the generalized Schr\"odinger equation at a given instant of $T$.

The assumptions that define the continuous case are quite general, the key requirement 
is the discreteness of the spectrum of each of the operators $\frac{1}{2}\ppi^2\hat{B}+\hat{C}_{\rm kin}$. 
The specific example of this case is the LQC Ashtekar-Paw{\l}owski-Singh model of a
FRW spacetime with negative cosmological constant \cite{bp-negL}. There the operator 
$\hat{B}^{-1}\hat{C}_{\rm kin}$ defined in the domain ${\cal D}_{gr}$ is essentially self adjoint 
in the Hilbert space obtained by introducing the new scalar product
$(\cdot\,|\,\hat{B}\cdot)_{\rm gr}$ in ${\cal H}_{\rm gr}$. That makes possible the 
Schr\"odinger-like construction of solutions to the quantum constraint (\ref{C}) relying on 
equation (\ref{3a}). That simpler construction gives a quantum theory equivalent to the one
derived in this section. Possibly, another class of examples (modulo, perhaps some easy 
generalization) can be found in case of a relativistic particle in a static spacetime.

It is worth noting, that in the spectral decomposition (\ref{cdec}) of the operator 
$\hat{C}$ there are degenerate points corresponding to the frequency $\ppi=0$ and the eigenvalues
$...<c_n(0)<...<c_1(0)= -c_0<0$ which however are located away from the neighborhood of $c=0$.
Nonetheless, we have also proposed an extension of the definition of each Hilbert space 
${\cal H}_{{\rm kin},c}$ to those points, by taking the limit $c\rightarrow c_n(0)$ of the 
functions $E_{\pm\ppi_n(c),c}$ and of their scalar product (see Remark at the end of Section 
\ref{Exp1prod}).

\section{The continuous case}\label{Exp2}

The departure point for our analysis is, similarly to the Section \ref{Exp1}, 
the framework specified via (\ref{Hkin}--\ref{psiqobspsi}). We apply it now to 
the case specified (up to technical assumptions) in the point \ref{it:case-cont} of 
Section \ref{sec:res}. The precise assumptions are specified below, in Section \ref{sec:cont-ass}.
Our goal is characterization of the solutions to the quantum constraint, as well 
as characterization of the relational quantum Dirac observables.

\subsection{Assumptions}\label{sec:cont-ass}

\begin{enumerate}
  \item In this section the Hilbert space is specified to be
    \be 
      {\cal H}_{\rm gr}\ =\ L^2(\mathbb{\mathbb{R}},\rd\nu_0) \ ,
    \ee
    where  $\nu_0$ is some measure, for example $\rd\nu_0(v)=\sum_{n\in\mathbb{N}}\delta(v-n)\rd v$.
  \item The domain ${\cal D}_{gr}\subset {\cal H}_{\rm gr}$ of the operators 
    $\hat{C}_{\rm gr}$, $\hat{B}$,  $\hat{B}^{-1}$, $\hat{C}_\ppi$ is the space of 
    smooth functions of a compact support and is preserved by the operators.\footnote{In 
      the case of the example with the measure $\rd\nu_0(v)=\sum_{n\in\mathbb{N}}\delta(v-n)\rd v$, 
      we drop the requirement of the smoothness in $v$.}
    $\hat{B}$ is bounded and positive. The action of the operators is extended by the 
    duality to every function $f:\mathbb{R}\rightarrow\mathbb{C}$ which defines a linear 
    functional ${\cal D}_{gr}\ni\psi\mapsto \int \rd\nu_0(v)\overline{f(v)}\psi(v)$. \label{it:cont-ass2}
  \item For every $\ppi\ge 0$, the operator $\hat{C}_\ppi$ (\ref{Cpi}) has absolutely continuous 
    spectrum $\mathbb{R}$, and its spectral decomposition consists of $1$-dimensional 
    Hilbert spaces. Furthermore, there exists a, normalized to the Dirac delta, basis 
    $\{{e}_{\ppi,c}\ :\ {c\in \mathbb{R}}\}$ of ${\cal H}_{\rm gr}$, where every 
    $e_{\ppi,c}:\mathbb{R}\rightarrow\mathbb{C}$ is an eigenfunction of the operator 
    $\hat{C}_\ppi$ with the eigenvalue $c$, that is
    \begin{subequations}\label{cprod3}\begin{align}
      \int \rd\nu_0(v) \overline{e_{\ppi,c}(v)}e_{\ppi,c'}(v)\ &=\ \delta(c-c') \ , &
      C_\ppi{e}_{\ppi,c}\ &=\ c {e}_{\ppi,c} \ . \tag{\ref{cprod3}}
    \end{align}\end{subequations}
    The eigenfunctions $e_{\ppi,c}$ are chosen in such a way that 
    $(\ppi,c)\mapsto e_{\ppi,c}(v)$ 
    is a continuous function at every $v\in\mathbb{R}$.
  \item There exits a function $a:\mathbb{R}^+\times \mathbb{R}\rightarrow\mathbb{R}$ 
    and a constant $b\in \mathbb{R}$ such that
    \be 
      \lim_{V\rightarrow\infty}\int_{-V}^V \rd\nu_0(v)
      \left(\overline{{e}_{\ppi,c}(v)}\hat{C}_{\rm gr}{e}_{\ppi',c}(v)\ -\
      \overline{\hat{C}_{\rm gr}{e}_{\ppi,c}(v)}{e}_{\ppi',c}(v)\right)\ 
      =\ b\sin(a(\ppi,c)-a(\ppi',c)) \ . \label{ecgre} 
    \ee
  \item The operator $\hat{B}$ extended to the space spanned by $e_{\ppi,c}$ satisfies
    \begin{align}
      \lim_{V\rightarrow\infty}\int_{-V}^V \rd\nu_0(v) 
      \left( \overline{e_{\ppi,c}(v)}\hat{B}e_{\ppi',c}(v)\ -\
      \overline{\hat{B}e_{\ppi,c}(v)}e_{\ppi',c}(v)\right)\ &=\ 0 \ ,\label{symB}\\
        \lim_{V\rightarrow \infty}\int_{-V}^V \rd v \overline{f(v)}\hat{B}f(v)\ &\ge 0 \ ,\\
    \end{align}
    and the map 
    \be 
      (\ppi,\ppi')\ \mapsto\ \lim_{V\rightarrow \infty}\int_{-V}^V \rd v 
      \overline{e_{\ppi,c}(v)}\hat{B}e_{\ppi',c}(v)
    \ee
    is well defined and continuous.
  \item The function $a$ and the function $c:(\ppi,c')\mapsto c'$ form a differentiable 
    coordinate system in  $\mathbb{R}^+\times\mathbb{R}$.
\end{enumerate}

\noindent{\bf Remark}
The condition (\ref{ecgre}) is a generalization of the symmetry of the operator 
$\hat{C}_{\rm gr}$ onto the space of the non-normalizable functions $e_{\ppi,c}$.

\subsection{The spectral decomposition of the constraint operator $\hat{C}$.}\label{Exm2decomp}

The first step toward solving the quantum constraint is deriving from the spectral 
decomposition (\ref{Hpic}) the decomposition (\ref{Hcpi}) of ${\cal H}_{\rm kin}$ 
corresponding to the operator $\hat{C}$ (\ref{C}). As a starting point we choose
the decomposition (\ref{Hpic}) of the operator $\hat{\Pi}$ followed by the decomposition of 
the operators $\hat{C}_\ppi$. From the assumptions listed in the section above it follows immediately 
that here the joint spectrum and the measure of the operators $\hat{\Pi}$ and ${\hat C}$ are
\be 
  {\rm Spec}\ =\ \mathbb{R}^2 \ ,\qquad \rd\ppi \rd\mu_\ppi(c)\ =\ \rd\ppi \rd c = \rd c \rd\ppi \ .
\ee 
Combining it with (\ref{psifurier}) and applying both to (\ref{Hpic}) we can express that
decomposition in the following, equivalent way
\be 
  \Psi(T)\ =\ \frac{1}{\sqrt{2\pi}}\int_{\mathbb{R}^2} \rd c \rd\ppi\,
  {\psi}_{\ppi,c}e^{i\ppi T} {e}_{|\ppi|,c} \ ,\label{PiCdecomp2}
\ee
where ${\cal H}_{\ppi c} = \mathbb{C}$. The Hilbert space
${\cal H}_{\rm kin}\ =\ L^2(\mathbb{R})\otimes{\cal H}_{\rm gr}$ is then unitarily related
with $L^2({\mathbb{R}^2})$ equipped with the measure $\rd\ppi \rd c$, that is
\begin{equation} 
  (\psi|\psi')_{\rm kin}\ =\ \int_{\mathbb{R}^2} \rd c \rd\ppi\,
  \overline{\psi_{\ppi,c}} \psi'_{\ppi,c} \ .\label{cprod3-kin}
\end{equation}
This completes the spectral decomposition of the operator $\hat{C}$.

\subsection{Solutions to the constraint $\hat{C}-c\hat{\id}$}\label{Exm2char}

Since (as already pointed out in the context of the discrete case) the measure $dc$ 
in (\ref{PiCdecomp2}) the Lebesgue one, the point $c=0$ of the spectrum of the operator $\hat{C}$ is of
measure $0$. Hence, as explained in Section \ref{Exp1}, we fix arbitrary value $c\in\mathbb{R}$ and 
consider the corresponding Hilbert space ${\cal H}_{{\rm kin},c}$ which can be identified with the 
solutions to the constraint $\hat{C}-c\hat{\id}$. We set $c=0$ only in a final step of the 
construction, first making sure that the result is stable with respect to the changes of $c$.

The Hilbert space ${\cal H}_{{\rm kin},c}$ defined by the
decomposition (\ref{PiCdecomp2}) is constructed out of the set
$\{E_{\ppi,c}\,|\,\ppi\in\mathbb{R}\}$ of functions of the variables
$(T,v)$
\be 
  E_{\ppi,c}(T,v)\ =\ \frac{1}{\sqrt{2\pi}}e^{i\ppi T}e_{|\ppi|,c}(v) \ ,
  \label{psi2}
\ee
and its scalar product $(\cdot|\cdot)_{{\rm kin},c}$ equals
\be 
  (E_{\ppi,c}|E_{\ppi',c})_{{\rm kin},c}\ =\ \delta(\ppi-\ppi') \ , 
\ee
where the Dirac delta is defined with respect to the Lebesgue measure $d\ppi$.
One can see, that this Hilbert space is unitarily related in a natural way 
with the Hilbert space of the formal integrals (see Section \ref{GAFRW})
\begin{equation} 
  \Psi\ =\ \int_{-\infty}^{\infty\oplus}\rd\ppi \Psi(\ppi)E_{\ppi,c}
  \label{intPsiE}
\end{equation}
with the scalar product
\be 
  (\Psi|\Psi')\ =\ \int_{\-\infty}^{\infty}\rd\ppi
  \overline{\Psi(\ppi)}\Psi'(\ppi) \ . \label{prodbasic} 
\ee
As in the discrete case, we can split the Hilbert space
\be 
  {\cal H}_{{\rm kin},c}\ =\ {\cal H}^+_{{\rm kin},c}\oplus {\cal H}^-_{{\rm kin},c} \ ,
  \label{HkincpmII}
\ee
where ${\cal H}^+_{{\rm kin},c}$ (${\cal H}^-_{{\rm kin},c}$) is spanned (in the 
integral sense) by the functions $E_{\ppi,c}$ of $\ppi\ge 0$ ($\ppi\le 0$).

In order to be able to view
the elements of ${\cal H}_{{\rm kin},c}$ as functions taking value in ${\cal H}_{\rm gr}$ 
which is endowed with a suitable new scalar product, let us consider the Hilbert 
space of the formal integrals
\be
  \psi\ =\ \int_{0}^{\infty\oplus} \rd\ppi \psi(\ppi)e_{\ppi,c}\label{formalint}
\ee
equipped with the scalar product
\be 
  (\psi|\psi')_c\ =\ \int_0^\infty \rd\ppi \overline{\psi(\ppi)}\psi'(\ppi) \ , \label{formalprod}
\ee
and denote the resulting Hilbert space by ${\cal H}_c$.
Then, every $\Psi^\pm\in {\cal H}^\pm_{{\rm kin},c}$ defines the map
\begin{equation}
  \mathbb{R}\ni\ T\mapsto\ \Psi^\pm(T)\in {\cal H}_c \ .
\end{equation}
For every $\Psi^\pm_1,\Psi_2^\pm\in {\cal H}^\pm_{{\rm kin},c}$,
the scalar product $(\Psi^\pm_1\,|\,\Psi_2^\pm)_{{\rm kin},c}$ can
be expressed by the ${\cal H}_c$ values
$\Psi^\pm_1(T),\Psi_2^\pm(T)$ at instant $T$ in the usual way
\be 
  (\Psi^\pm_1\,|\,\Psi_2^\pm)_{{\rm kin},c}\ 
  =\ (\Psi^\pm_1(T)\,|\,\Psi_2^\pm(T))_{c} \ , 
\ee
where the right hand side is independent of the choice of the value
of the $T$ variable.\\

\noindent{\bf Remark:}
The construction in this subsection was performed in a way similar to the discrete 
case described in Section \ref{Exp1}. There is however a significant
difference in comparison to that case: the formal integrals should not be interpreted 
as the actual integrals. In particular, the formal integral (\ref{intPsiE}) should not be 
identified with a function
\be 
  (T,v)\ \mapsto\ \int_{-\infty}^{\infty} \rd\ppi \Psi(\ppi)E_{\ppi,c}(T,v) \ ,
\ee
even when a latter one is well defined. Also an element $\psi\in {\cal H}_c$
can not be identified  with a function
\be 
  v\ \mapsto\ \int_0^\infty \rd\ppi \psi(\ppi)e_{\ppi,c}(v) \ .
\ee
The reason for this lack of the correspondence will become clear in the next subsection.

\subsection{The scalar product between the solutions}\label{Exm2prod}

The scalar product $(\cdot|\cdot)_{c}$ in the space of formal integrals (\ref{formalint}) 
has been introduced in the previous subsection just by a declaration -- (\ref{formalprod}). 
Below we will show, that this product can be interpreted in terms of the Hilbert
space ${\cal H}_{\rm gr}$ similarly to the discrete case. The difference in 
the structure of the spectrum of $\hat{C}$ induces here however an important difference 
with respect to the former case.

Our starting point in finding the relation is the assumed property
\be
  \lim_{V\rightarrow\infty}\int_{-V}^V \rd v\,
  \left(\overline{{e}_{\ppi,c}(v)}\hat{C}_{\rm gr}{e}_{\ppi',c}(v)\ -\
  \overline{\hat{C}_{\rm gr}{e}_{\ppi,c}(v)}{e}_{\ppi',c}(v)\right)\ 
  =\ b\sin(a(\ppi,c)-a(\ppi',c)) \ ,
\ee
combined with the extended symmetry (\ref{symB}) of the operator $\hat{B}$ and the 
asymptotic properties of the functions $e_{\ppi,c}$ given by 
\be 
  \hat{C}_{\rm gr}e_{\ppi,c}\ =\ (c\hat{\id}-\frac{1}{2}\ppi^2\hat{B})e_{\ppi,c} \ .
\ee
Applying the above elements we arrive to the equality
\begin{equation}
  \lim_{V\rightarrow \infty} \int_{-V}^V \rd v\, \overline{{e}_{\ppi,c}(v)} \hat{B}
  {e}_{\ppi',c}(v)\ =\ 2b\frac{\sin(a(\ppi,c)-a(\ppi',c))}{\ppi^2-\ppi'^2} \ .
  \label{limebe'}
\end{equation}
Taking the limit $\ppi'\rightarrow \ppi$, we obtain
\begin{equation}
  \int_{-\infty}^\infty \rd v\, \overline{{e}_{\ppi,c}(v)} \hat{B}
  {e}_{\ppi,c}(v)\ =\ \frac{b}{\ppi}\frac{\partial a(\ppi,c)}{\partial \ppi} \ .\label{eBe}
\end{equation}
This result means in particular, that if we introduce in the Hilbert space 
${\cal H}_{\rm gr}$ a new scalar product $(\cdot\,|\,\cdot)_{\hat{B}}$ and take the 
completion, or equivalently, in the space of functions defined on $\mathbb{R}$ introduce 
the following scalar product
\begin{equation}\label{eq:condBprod-def}
  (f|g)_{\hat{B}}\ :=\ \int_{-\infty}^{\infty} \rd v \overline{f(v)}\hat{B}g(v) \ ,
\end{equation}
then the function $e_{\ppi,c}$ is normalizable (with respect to it) for every 
$c$ and every $\ppi> 0$. (The point $\ppi=0$ is of the spectral decomposition measure zero and hence not relevant.) 

Finally, comparing \eqref{limebe'} with \eqref{eq:condBprod-def} we arrive to the scalar product for two different $\ppi,\ppi'$,
\begin{equation}
  (e_{\ppi,c}\,|\,e_{\ppi',c})_{\hat{B}}\ 
  =\ 2b\frac{\sin(a(\ppi,c)-a(\ppi',c))}{\ppi^2-\ppi'^2} \ .\label{epiBepi'}
\end{equation}

In general, the right hand side is not zero. However, given a  value $a'$ of the 
function $a$ (and given in this subsection $c\in\mathbb{R}$) there is a distinguished 
set of the values taken by the label $\ppi$ labelling the functions $e_{\ppi,c}$, namely
\begin{equation}
  \mathbb{R}_{c,a'}\ =\ \{\ppi\in\re^+
  :\ (a(\ppi,c)-a')\ \in\ \pi\mathbb{Z}\} \ .
\end{equation}
The corresponding functions
\begin{equation}
  \{e_{\ppi,c}\,:\,\ppi\in \mathbb{R}_{c,a'}\}
\end{equation}
are orthogonal to each other, specifically 
\begin{equation}
  (e_{\ppi,c}\, | \,e_{\ppi',c})_{\hat{B}}\ =\
  \frac{b}{\ppi}\frac{\partial a(\ppi,c)}{\partial\ppi}\delta_{\ppi,\ppi'}, \ \label{ebe2}
\end{equation}
where $\delta_{\cdot,\cdot}$ is the Kronecker delta. It is sufficient to restrict 
the values $a'$ in the definition of $\mathbb{R}_{c,a'}$ to the interval $a'\in[0,\pi)$.
The properties of the functions $e_{\ppi,c}$ justify Remark made at the end of the previous 
subsection, which pointed out differences between current continuous case and 
the discrete one studied in the previous section. To derive a representation of the 
Hilbert space ${\cal H}_c$ analogous to that of Section \ref{Exp1} we go back to the 
formal integral (\ref{formalint}) and the scalar product $(\cdot|\cdot)_c$ (\ref{formalprod}), 
and change the variable of the integration for the function $a$.
This way we get
\begin{subequations}\begin{align}
  \begin{split}
    \psi\ =\ \int_{0}^{\infty\oplus} \rd\ppi \psi(\ppi)e_{\ppi,c}\ 
    &=\ \int_{[0,\pi)}^\oplus \rd a \sum_{\ppi\in \mathbb{R}_{c,a}}
    \left(\left|\frac{\partial a(\ppi,c)}{\partial \ppi}\right|^{-\frac{1}{2}}\psi(\ppi)\right)
    \left(\left|\frac{\partial a(\ppi,c)}{\partial \ppi}\right|^{-\frac{1}{2}}e_{\ppi,c}\right)\  \\ 
    &=\ \int_{[0,\pi)}^{\oplus} \rd a 
    \sum_{\ppi\in \mathbb{R}_{c,a}}\check{\psi}(\ppi)\check{e}_{\ppi,c} \ ,
  \end{split} \label{intdasum}\\
  (\psi|\psi')_c\ &=\ \int_{[0,\pi)}^{\oplus} \rd a 
  \sum_{\ppi\in \mathbb{R}_{c,a}}\overline{\check{\psi}(\ppi)}\check{\psi}'(\ppi)
  \ ,
\end{align}\end{subequations}
where $\check{\psi}(\ppi)$ and, respectively, $\check{e}_{\ppi,c}$ are defined by 
the factors in the paratheses in the $1$st line of \eqref{intdasum}.

The scalar product $(\cdot|\cdot)_c$ can be  expressed by the product $(\cdot|\cdot)_{\hat{B}}$
via the relation (\ref{ebe2})
\begin{equation}
  (\psi|\psi')_c\ =\ \int_{[0,\pi)} \frac{\rd a}{b} 
  (\sum_{\ppi\in \mathbb{R}_{c,a}}\check{\psi}(\ppi)
  \check{e}_{\ppi,c}\,|\,\sum_{\ppi'\in \mathbb{R}_{c,a}}\ppi'\check{\psi}'(\ppi')\check{e}_{\ppi',c})_{\hat{B}}
  \ .
\end{equation}

The emerging structure can be described as follows. For every $a\in[0,\pi)$ introduce 
the vector space ${\rm Span}(e_{\ppi,c}\,|\,\ppi\in \mathbb{R}_{c,a})$ and endow it with 
an operator
\begin{equation}
  \hat{\Pi}_{c,a}e_{\ppi,c}\ :=\ \ppi e_{\ppi,c}
\end{equation}
and with the scalar product
\begin{equation}
  (\psi|\psi')_{c,a}\ :=\ (\psi|\hat{\Pi}_{c,a}\psi')_{\hat{B}} \ .\label{productca}
\end{equation}
Denote the resulting Hilbert space by ${\cal H}_c^{(a)}$. Then every formal 
integral (\ref{intdasum}) becomes an integral of vectors $\psi^{(c,a)}$
\begin{subequations}\label{eq:cont-descr}\begin{align}
  \psi\ &=\ \int^\oplus_{[0,\pi)}\psi^{(c,a)} \ , &
  \psi^{(c,a)}\ &
  =\ \sum_{\ppi\in \mathbb{R}_{c,a}}\check{\psi}(\ppi)\check{e}_{\ppi,c}\ \in\ {\cal H}_c^{(a)}
\tag{\ref{eq:cont-descr}}\end{align}\end{subequations}
and the scalar product reads
\begin{equation}
  (\psi\,|\,\psi')_c\ =\ \int_{[0,\pi)} \frac{\rd a}{b} (\psi^{(c,a)}|\psi'^{(c,a)})_{c,a} \ .
\end{equation}
As in the previous section we note, that the operator $\hat{\Pi}_{c,a}$ can be expressed 
in terms of the operators $\hat{C}_{\rm gr}$ and $\hat{B}$ via
\begin{equation}
  \hat{\Pi}_{c,a}e_{\ppi,c}\ =\ \sqrt{2\hat{B}^{-1}(c\hat{\id}-\hat{C}_{\rm gr})}\,e_{\ppi,c} \ ,\quad 
  \ppi\in \mathbb{R}_{c,a} \ .
\end{equation}

\subsection{The Dirac observables}\label{sec:cont-dirac}

Let us now focus on the construction of the Dirac observables. As in the discrete case 
the operator $\hat{\Pi}\otimes \hat{\id}$ defined in 
${\cal H}_{\rm kin}=L^2(\mathbb{R})\otimes{\cal H}_{\rm kin}$ is a quantum Dirac observable 
as well as any $f(\hat{\Pi})\otimes \hat{\id}$.

Here we will consider the class of the Dirac observables constructed (via the very 
same technique as the one used in Section \ref{diracc1}) by the relational 
observable  method (see (\ref{qobsC})) from all the operators
\be 
  \hat{F}\ =\ \hat{\id}\otimes\hat{G} \ ,
\ee
where $\hat{G}$ is an operator in ${\cal H}_{\rm gr}$.

The general formula for the observable (\ref{qobsC}) reads
\be 
  \hat{F}^{D}_{(T,t)}\ =\ \int \rd\tau e^{-i\tau \hat{C}}\circ {\rm Sym} \left(\,|\,\hat{\Pi}\otimes \hat{B}
 \,|\,\circ \hat{F}\circ \delta(\hat{T}-t\hat{\id})\otimes \hat{\id}\right)\circ e^{i\tau\hat{C}} \ ,
\ee
where the symmetrization ${\rm Sym}$ is the one defined already in Section \ref{diracc1}. 
Since in the discrete case it leads to clear and physically reasonable results we apply
it also here without change.
Given this, in the case of the kinematical observable $\hat{F}=\hat{\id}\otimes\hat{G}$, the
relational quantum Dirac observable $\hat{G}^D_{(T,t)}$ is given by
\be 
  \hat{G}^D_{(T,t)}\ =\ \int \rd\tau e^{-i\tau \hat{C}} \left(
    \theta(\hat{\Pi})|\hat{\Pi}|^{\frac{1}{2}}
    \delta(\hat{T}-t\hat{\id})|\hat{\Pi}|^{\frac{1}{2}} \theta(\hat{\Pi}) +
    \theta(-\hat{\Pi})|\hat{\Pi}|^{\frac{1}{2}}
    \delta(\hat{T}-t\hat{\id})|\hat{\Pi}|^{\frac{1}{2}} \theta(-\hat{\Pi}) \right)
  \otimes \sqrt{B}\hat{G}\sqrt{B} e^{i\tau\hat{C}} \ .
\ee
In terms of the spectral decomposition (\ref{PiCdecomp2}), the sesquilinear form 
$\hat{G}^D_{(T,t)}$ in ${\cal H}_{\rm kin}$ reads
\begin{equation} \begin{split}
  (\psi\,|\,\hat{G}^D_{(T,t)}\psi')_{\rm kin}\ 
  =\ \int_{-\infty}^\infty \rd c&\int_{0}^\infty \rd\ppi\int _0^\infty 
  \rd\ppi'(e^{-it(\ppi-\ppi')}\overline{\psi_{\ppi,c}}\psi'_{\ppi',c}+\
  e^{it(\ppi-\ppi')}\overline{\psi_{-\ppi,c}}\psi'_{-\ppi',c})\cdot\ \\
  &\cdot\int \rd v \overline{e_{\ppi,c}(v)} \hat{B}^{\frac{1}{2}}\hat{G}\hat{B}^{\frac{1}{2}}\,
  e_{\ppi',c}(v)\sqrt{\ppi\ppi'}\
\end{split}\label{GDsesqui2}\end{equation}
and induces a sesquilinear form of the space of solutions to the constraint $\hat{C}-c\hat{\id}$. 
Indeed, given two solutions
\begin{subequations}\begin{align}
  \psi(T)\ &=\ \frac{1}{\sqrt{2\pi}}\int_0^{\infty\oplus} \rd\ppi(\psi^+(\ppi) 
  e^{i\ppi T}e_{\ppi,c}\ +\ \psi^-(\ppi) e^{-i\ppi T}e_{\ppi,c})  \ , \\
  \psi'(T)\ &=\ \frac{1}{\sqrt{2\pi}}\int_0^{\infty\oplus} \rd\ppi'(\psi'^+(\ppi') 
  e^{i\ppi' T}e_{\ppi',c}\ +\ \psi'^-(\ppi') e^{-i\ppi' T}e_{\ppi',c}) \ ,
\end{align}\end{subequations}
we have
\begin{equation}\begin{split}
  (\psi\,|\,\hat{G}^D_{(T,t)c}\psi')_{{\rm kin},c}\ 
  =\ \int_{0}^\infty \rd\ppi\int _0^\infty &\rd\ppi'\sqrt{\ppi\ppi'}
  \int \rd v \overline{e_{\ppi,c}(v)}
  \hat{B}^{\frac{1}{2}}\hat{G}\hat{B}^{\frac{1}{2}}\,
  e_{\ppi',c}(v) \cdot  \\
  &(e^{-it(\ppi-\ppi')}\overline{\psi^+(\ppi)}\psi'^+(\ppi')\
  +\ e^{it(\ppi-\ppi')}\overline{\psi^-(\ppi)}\psi'^-(\ppi') \ .
\end{split}\label{GDsesqui3}\end{equation}
Finally, we can apply the decomposition into the subspaces ${\cal H}^{(a)}_{c}$, obtaining
\begin{equation}\begin{split}
  (\psi\,|\,\hat{G}^D_{(T,t)c}\psi')_{{\rm kin},c}\ 
  =\ \int_{[0,\pi)} \rd a&\int _{[0,\pi)} \rd a' \sum_{\ppi\in \mathbb{R}_{c,a}}\sum_{\ppi'\in \mathbb{R}_{c,a'}}
  \sqrt{\ppi\ppi'} ({\check{e}_{\ppi,c}}\,|\, \hat{B}^{-\frac{1}{2}}\hat{G}\hat{B}^{\frac{1}{2}}\,
  \check{e}_{\ppi',c})_{\hat{B}}\cdot  \\
  &\left( e^{-it(\ppi-\ppi')}\overline{\check{\psi}^+(\ppi)}\check{\psi}'^+(\ppi')\
  +\ e^{it(\ppi-\ppi')}\overline{\check{\psi}^-(\ppi)}\check{\psi}'^-(\ppi')\right) \ .
\end{split}\label{psiGpsi'}\end{equation}
This formula immediately implies the quite peculiar property of defined observable, namely
the presence of the cross terms $a\not= a'$, which breaks the diagonality of the 
integral on the right hand side. That property makes even the (supposedly) trivial 
case of $\hat{G}=\hat{\id}$, quite non-trivial. Indeed, the  relational Dirac observable
$\hat{1}^D_{(T,t)}$ corresponding to it is given by
\begin{equation}\begin{split}
  (\psi\,|\,\hat{1}^D_{(T,t)}\psi')_{\rm kin}\ 
  =\ \int_{-\infty}^{\infty} \rd c\int_{0}^\infty &\rd\ppi\int _0^\infty \rd\ppi'\sqrt{\ppi\ppi'}
  ({e_{\ppi,c}(v)}\,|e_{\ppi',c})_{\hat{B}}\cdot  \\
  &(e^{-it(\ppi-\ppi')}\overline{\psi_{\ppi,c}}\psi'_{\ppi',c}\
  +\ e^{it(\ppi-\ppi')}\overline{\psi_{-\ppi,c}}\psi'_{-\ppi',c}) \ .
\end{split}\label{GDsesqui2-id}\end{equation}
However, the right hand side is not the identity because
\begin{equation}
  \sqrt{\ppi\ppi'}({e_{\ppi,c}(v)}\,|e_{\ppi',c})_{\hat{B}}\ \neq\ \delta(\ppi-\ppi') \ ,
\end{equation}
and instead is given by (\ref{epiBepi'}). The classical origin/meaning of this property 
was explained in Section \ref{GAFRW}.

\subsection{The constraint as evolution  in ${\cal H}_{\rm gr}$}\label{Exp2evolution}

In the discrete case it was possible to view the quantum  constraint defined by 
the operator $\hat{C}-c\hat{\id}$ as a unitary evolution (\ref{sol2}) in the Hilbert space 
${\cal H}_{\rm gr}$ of the kinematical degrees of freedom of the quantum geometry. Here 
we derive an analogous characterization for the solutions in the continuous case.

In order to do so we recall from Subsections \ref{Exm2char} and \ref{Exm2prod}, that
the solutions to the quantum constraint defined by the operator $\hat{C}-c\hat{\id}$ form 
the Hilbert space ${\cal H}^-_{{\rm kin},c}\oplus{\cal H}^+_{{\rm kin},c}$ (\ref{HkincpmII}), 
and the subspaces ${\cal H}^\pm_{{\rm kin},c}$ consist of the formal integrals
\be 
  \int_0^{\infty\oplus} \rd\ppi \psi(\ppi)E_{\pm\ppi,c}
\ee
of the functions (\ref{psi2}).
Furthermore, each element $\psi^\pm\in{\cal H}^\pm_{{\rm kin},c}$ can be viewed 
as a function
\begin{equation}
  \mathbb{R}\ni T\mapsto \psi^\pm(T)\ =\ \int_{[0,\pi)}^{\oplus} \rd a \psi^{(c,a)}(T)\ \in\
  \int_{[0,\pi)}^{\oplus}\rd a {\cal H}_c^{(a)} \ ,\label{psipm(T)}
\end{equation}
where the family of the Hilbert spaces ${\cal H}_c^{(a)}$ is defined by the scalar product 
(\ref{productca}) introduced  in the vector space ${\rm Span}(e_{\ppi,c}\,|\,\ppi\in \mathbb{R}_{c,a})$.
This means, that each solution $\psi^\pm$ is a family  of functions labelled by $a\in [0,\pi)$
\begin{equation}
  \mathbb{R}\ni T\mapsto \psi^{\pm(c,a)}(T)\ \in\ {\cal H}_c^{(a)} \ , \label{psiacT}
\end{equation}
where the scalar product between two solutions equals
\begin{equation}
  (\psi^\pm\,|\,\psi'^\pm)\ =\ \int_{[0,\pi)} \frac{\rd a}{b} (\psi^{\pm(c,a)}(T)\,|\,\psi'^{\pm(c,a)}(T))_{c,a} \ ,
\end{equation}
with the right hand side of this equality being independent of $T$.

In each of the Hilbert spaces ${\cal H}^{(a)}_c$, the operator ${2\hat{B}^{-1}(c\hat{\id}-\hat{C}_{\rm gr})}$  
is defined on the domain ${\rm Span}(e_{\ppi,c}\,:\ppi\in \mathbb{R}_{c,a})$ and becomes 
an essentially self-adjoint, positive operator ${2\hat{B}^{-1}(c\hat{\id}-\hat{C}_{\rm gr})}_{(c,a)}$. 
(We will go back to the global definiteness of this and related operators at the end of this description.) 
With use of this operator, every function (\ref{psiacT}) can be written in the form
\begin{equation}
  \psi^{\pm(c,a)}(T)\ 
  =\ e^{\pm i T\sqrt{{2\hat{B}^{-1}(c\hat{\id}-\hat{C}_{\rm gr})}_{(c,a)} }}\,\psi^{\pm(c,a)}(0) \ .\label{evolution2a}
\end{equation}

For each of the Hilbert spaces ${\cal H}_c^{(a)}$, $a\in[0,\pi)$, there is a naturally 
defined unitary embedding into the Hilbert space ${\cal H}_{\rm gr}$
\be 
  \hat{B}^{\frac{1}{2}}(2\hat{B}^{-1}(c\hat{\id}-\hat{C}_{\rm gr})_{c,a})^{\frac{1}{4}}\ 
  :\ {\cal H}_c^{(a)}\ \rightarrow {\cal H}_{\rm gr} \ .
  \label{unitaryac}
\ee
It maps (\ref{evolution2a}) into
\be 
  \tilde{\psi}^{\pm(c,a)}(T)\ =\ e^{\pm i T \sqrt{{2\hat{B}^{-\frac{1}{2}}(c\hat{\id}-C_{\rm gr})
  \hat{B}^{-\frac{1}{2}}}_{(c,a)}}}\,\tilde{\psi}^{\pm(c,a)}(0)\label{schroed2} \ , 
\ee
where  the operator ${2\hat{B}^{-1}(c\hat{\id}-\hat{C}_{\rm gr})}_{(c,a)}$ is mapped into 
the operator ${2\hat{B}^{-\frac{1}{2}}(c\hat{\id}-\hat{C}_{\rm gr})\hat{B}^{-\frac{1}{2}}}_{(c,a)}$ defined 
in the domain
\be 
  {\rm Span}(\hat{B}^{\frac{1}{2}}e_{\ppi,c}\,|\,\ppi\in \mathbb{R}_{c,a}) \ . 
\ee
The image of the map ${\cal H}^{(a)}_{c}\rightarrow {\cal H}_{\rm gr}$ (\ref{unitaryac}), 
denoted here by $\tilde{\cal H}^{(a)}_{c}$, is the completion of 
${\rm Span}({\hat{B}^{\frac{1}{2}}}e_{\ppi,c}\,|\,\ppi\in \mathbb{R}_{c,a})\subset{\cal H}_{\rm gr}$, 
and is a proper subspace of ${\cal H}_{\rm gr}$. For two different $a\not= a'$, the corresponding 
subspaces satisfy
\be 
  \tilde{\cal H}^{(a)}_{c}\ \not=\ \tilde{\cal H}^{(a')}_{c} \ , \qquad
  \tilde{\cal H}^{(a)}_{c}\ \not\bot\ \tilde{\cal H}^{(a')}_{c} \ .
\ee

Let us discuss now the
definitions of the operators $2\hat{B}^{-1}(c\hat{\id}-\hat{C}_{\rm gr})_{(c,a)}$, 
$2\hat{B}^{-\frac{1}{2}}(c\hat{\id}-\hat{C}_{\rm gr})\hat{B}^{-\frac{1}{2}}_{(c,a)}$ as well as the 
operators $2\hat{B}^{-1}(c\hat{\id}-\hat{C}_{\rm gr})$,
$2\hat{B}^{-\frac{1}{2}}(c\hat{\id}-\hat{C}_{\rm gr})\hat{B}^{-\frac{1}{2}}$. For that
we employ the Assumption \ref{it:cont-ass2}. Each of the operators ${\hat B}$, ${\hat B}^{-1}$, 
$\hat{C}_{\rm gr}$ is defined by the extension of the corresponding operator defined originally 
in the domain ${\cal D}_{gr}$, onto the space ${\cal D}^*_{gr}$ of functions $f:\mathbb{R}\rightarrow \mathbb{C}$ 
dual to the domain ${\cal D}_{gr}$ in the sense of the measure $\nu_0$. Moreover, each of the extended 
operators preserves ${\cal D}^*_{gr}$, because the original operators preserve ${\cal D}_{gr}$.
The composition of the extended operators defines the operator
$\hat{B}^{-1}(c\hat{\id}-\hat{C}_{\rm gr})$, and each function
$e_{\ppi,c}$ is the eigenfunction of this operator corresponding to the eigenvalue $\ppi^2$. 
The restriction of the operator $2\hat{B}^{-1}(c\hat{\id}-\hat{C}_{\rm gr})$ to the vector space 
${\rm Span}(e_{\ppi,c}\,:\, \ppi\in\mathbb{R}_{(c,a)})$, given $(c,a)$,
defines a self adjoint, positive operator in the Hilbert space ${\cal H}^{(a)}_c$ which has 
a well defined square root. The operator  $2\hat{B}^{-1}(c\hat{\id}-\hat{C}_{\rm gr})$ defined in 
${\cal D}^*_{gr}$ admits (by the restriction) an action in a subspace of the Hilbert space 
${\cal H}_{{\rm gr},\hat{B}}$ defined by introducing in ${\cal H}_{\rm gr}$ the new Hilbert 
product $(\cdot\,|\hat{B}\,\cdot)$ and taking the completion. This restriction coincides with 
the operator $2\hat{B}^{-1}(c\hat{\id}-\hat{C}_{\rm gr})^\dagger$ adjoint in  
${\cal H}_{{\rm gr},\hat{B}}$ to $2\hat{B}^{-1}(c\hat{\id}-\hat{C}_{\rm gr})$
considered in the domain ${\cal D}_{gr}$. But this operator is not symmetric and its square root 
is not well defined either. Each operator 
$2\hat{B}^{-\frac{1}{2}}(c\hat{\id}-C_{\rm gr})\hat{B}^{-\frac{1}{2}}_{(c,a)}$ is defined just as the
transformation of $2\hat{B}^{-1}(c-\hat{C}_{\rm gr})_{(c,a)}$ by the map (\ref{unitaryac}). 
A single operator in ${\cal H}_{\rm gr}$, whose restriction to 
${\rm Span}(\hat{B}^{\frac{1}{2}}e_{\ppi,c}\,:\,\ppi\in\mathbb{R}_{c,a})$
is $2\hat{B}^{-\frac{1}{2}}(c\hat{\id}-\hat{C}_{\rm gr})\hat{B}^{-\frac{1}{2}}_{(c,a)}$ can be defined as the
transformation of the operator $2\hat{B}^{-1}(c\hat{\id}-\hat{C}_{\rm gr})^\dagger$ by (\ref{unitaryac}). 
The result can be defined in the equivalent way as follows. In the domain 
$\hat{B}^{\frac{1}{2}}{\cal D}_{gr}\subset {\cal H}_{\rm gr}$  consider the operator $2\hat{B}^{-\frac{1}{2}}(c\hat{\id}-\hat{C}_{\rm gr})\hat{B}^{-\frac{1}{2}}$. 
The adjoint $[2\hat{B}^{-\frac{1}{2}}(c\hat{\id}-\hat{C}_{\rm gr})\hat{B}^{-\frac{1}{2}}]^\dagger$ 
is then the operator we seek.

In summary, given $c\in \mathbb{R}$, we defined a family
$(\tilde{\cal H}_c^{(a)})_{a\in[0,\pi)}$ of subspaces of ${\cal H}_{\rm gr}$. 
In ${\cal H}_{\rm gr}$ the operator $[2\hat{B}^{-\frac{1}{2}}(c\hat{\id}-C_{\rm
gr})\hat{B}^{-\frac{1}{2}}]^\dagger$ is well defined, but not symmetric.
However its restriction to each of the spaces
\begin{equation}
  {\cal D}_c^{(a)}\ :=\ {\rm Span}({\hat{B}^{\frac{1}{2}}}e_{\ppi,c}\,|\,\ppi\in \mathbb{R}_{c,a})
\end{equation}
defines an essentially-self adjoint and positive operator in the corresponding completion
$\tilde{\cal H}_c^{(a)}\subset {\cal H}_{\rm gr}$. Every solution (\ref{evolution2a}) 
is mapped by (\ref{unitaryac}) into $\tilde{\psi}$,
\begin{equation}
  \mathbb{R}\ni T\mapsto \tilde{\psi}^\pm(T)\ =\ \int_{[0,\pi)}^{\oplus} \rd a \tilde{\psi}^{(c,a)}(T)\ \in\
  \int_{[0,\pi)}^{\oplus}\rd a \tilde{\cal H}^{(a)}_c \ ,\label{checkpsipm(T)}
\end{equation}
where each component $\tilde{\psi}^{\pm(c,a)}(T)$ is defined by (\ref{schroed2})
and the scalar product between two  solutions ${\psi}^\pm,{\psi}'^\pm\in$ is
\begin{equation}
  (\psi^\pm |\psi'^\pm)_{{\rm kin},c}\ 
  =\ \int_{[0,\pi)} \rd a (\tilde{\psi}^{\pm(c,a))}(T)\,|\,\tilde{\psi}'^{\pm(c,a))}(T))_{\rm gr} \ .
\end{equation}

Up to this point, we could say that every solution to the quantum constraint in 
the continuous case can be viewed as a family of solutions similar to those encountered 
in the discrete one, with the extra integral with respect to $a$ in the scalar product. 
In particular, the evolution (\ref{schroed2}) defined by the constraint $\hat{C}-c\hat{\id}$ 
in the case considered here reduces to each of the subspaces $\tilde{\cal H}_c^{(a)}$, 
independently of the others.

However, as if was shown in Section \ref{sec:cont-dirac}, the relational Dirac observables 
break the diagonality of this picture, since they have non-zero cross terms between 
two different spaces $\tilde{\cal H}_c^{(a)}$ and $\tilde{\cal H}_c^{(a')}$. 
Indeed, in terms of the formula (\ref{checkpsipm(T)}), given two solutions ${\psi}$ and ${\psi'}$ 
to the quantum constraint $\hat{C}-c\hat{\id}$ and the corresponding transformed functions 
$\tilde{\psi}$ and $\tilde{\psi}'$, the Dirac observable (\ref{psiGpsi'}) defined by 
a kinematical observable $\hat{G}$ in ${\cal H}_{\rm gr}$ takes the following form
\begin{equation}
  (\psi\,|\,\hat{G}^D_{(T,t)c}\psi')_{{\rm kin},c} \ =\ \frac{1}{b^2}\int_{[0,\pi)} \rd a\int_{[0,\pi)}\rd a' 
  \left(\,\check{\psi}^{\pm(c,a)}(t)\,|\,\hat{G}\check{\psi}'^{\pm(c,a')}(t)\,\right)_{\rm gr} \ .\label{GDsesqui4}
\end{equation}

Also the map $\hat{\Pi}\otimes \hat{\id}\ \mapsto\ \hat{\Pi}^{\rm D}_{(T,t)}$ is no longer the identity, namely
\begin{equation} 
  (\psi\,|\, \hat{\Pi}^{\rm D}_{(T,t)}\psi')_{{\rm kin},c}\ 
  =\ \frac{1}{b^2}\int_{[0,\pi)^2} \rd a \rd a'(\tilde{\psi}^{(c,a)}
  \,|\,\tilde{\Pi}_{c,a'}\tilde{\psi}^{(c,a')})_{\rm gr} \ .
\end{equation}

\subsection{Discussion, the limit $c\rightarrow 0$}

In the previous subsections we have described, for arbitrarily fixed value of $c$, the 
corresponding Hilbert space ${\cal H}_{{\rm kin},c}$ in the spectral decomposition of the 
operator $\hat{C}$ as well as we have introduced the relational quantum Dirac observables therein.

As in the discrete case, the ambiguity in the definition of the functions $E_{\ppi,c}$,
\be
  E_{\ppi,c}\ \mapsto\ e^{i\alpha(\ppi,c)}E_{\ppi,c},
\ee
does not affect the definition of ${\cal H}_{{\rm kin},c}$ nor any other structure 
we introduced to characterize it.

The continuity of the map 
\be 
  c\ \mapsto\ E_{\ppi,c}
\ee
provided by assumption 3, ensures the continuity in $c$ of the Hilbert space ${\cal H}_{{\rm kin},c}$, therefore the Hilbert 
space ${\cal H}_{{\rm kin},0}$ is uniquely defined, as well as are the relational Dirac 
observables therein. There is also the continuity in $c$ of the structures we have introduced 
to characterize the resulting quantum theory, namely
\begin{enumerate}[(i)]
  \item $\int_0^{\oplus\pi} \rd a\tilde{\cal H}^{(a)}_c\ =\ \overline{{\rm Span}(\hat{B}^{\frac{1}{2}}
    e_{\ppi,c}\,|\, \ppi\in\mathbb{R}_{c,a})}\ \subset\ {\cal H}_{\rm gr}$, $a\in [0,\pi)$,
  \item
    $\tilde{\Pi}_{c,a}\ 
    =\ \sqrt{\hat{B}^{-\frac{1}{2}}(c\hat{\id}-\hat{C}_{\rm gr}) \hat{B}^{-\frac{1}{2}}_{(c,a)}}$.
\end{enumerate}
Although there exist degenerate points in the definition of individual Hilbert spaces 
${\cal H}^{(a)}_{c}$ similar to those in the discrete case, the integral along $[0,\pi)$
smooths them out.

The resulting quantum theory is defined independently in two copies of the formal integral 
$\int_0^{\oplus\pi} \rd a{\cal H}^{(a)}_0$ of a family of subspaces 
$\tilde{\cal H}^{(a)}_{0}\subset{\cal H}_{\rm gr}$. In each copy the dynamics if defined
by a family of the Hamiltonian operators 
$(\pm\sqrt{-\hat{B}^{-\frac{1}{2}}(\hat{C}_{\rm gr}) \hat{B}^{-\frac{1}{2}}_{(0,a)}})_{a\in[0,\pi)}$. 
Whereas the dynamics in each term ${\cal H}^{(a)}_c$ of the formal integral is defined independently 
of the other terms, the relational Dirac observables mix different terms ${\cal H}^{(a)}_c$.

An explicit example of Case II is the LQC Ashtekar-Paw{\l}owski-Singh model of the FRW 
spacetime with the positive cosmological constant \cite{ap-posL}. In that case the analysis 
based on the Schr\"odinger picture (\ref{3a}), which uses the
operator $\hat{B}^{-1}\hat{C}_{\rm gr}$ as the evolution one, faces a technical problem: 
the operator is not essentially self adjoint and admits inequivalent
self-adjoint extensions. In fact, the 1-dimensional family of the spaces 
${\rm Span}(e_{\ppi,c}\,|\, a(\ppi,c)-a'=\pi n,\ n\in\mathbb{Z})$ we introduce for every 
$a'\in[0,\pi)$, corresponds exactly to the 1-dimensional family of the inequivalent self-adjoint
extensions of $\hat{B}^{-1}\hat{C}_{\rm gr}$. The approach (\ref{3a}) gives a result 
essentially different than the one found here: it forces us to choose one of the self-adjoint 
extensions. The resulting theory is then formed by two copies (the positive/negative frequencies) of the
Schr\"odinger like quantum mechanics defined in the subspace ${\cal H}^{(a)}_0\subset {\cal H}_{\rm gr}$ 
corresponding to an arbitrarily fixed $a\in[0,\pi)$. The Hamiltonian operator is
$\pm\sqrt{-\hat{B}^{-\frac{1}{2}}(\hat{C}_{\rm gr}) \hat{B}^{-\frac{1}{2}}_{(0,a)}}$ 
and the relational observables do not have any option to mix two different subspaces ${\cal H}^{(a)}_0$ 
and ${\cal H}^{(a')}_0$, for $a\not= a'$. In consequence, by construction, we have there
\be 
  \hat{1}^{\rm D}_{(T,t)} \ =\ \hat{\id} \ .
\ee
From the point of view of the known, corresponding classical theory, for as long as we 
prescribe the fundamental role in describing the evolution to the internal time, that result is
incorrect, because the relational observable $1^{\rm D}_{(T,t)}$ constructed from the constant 
function $1$ is either $1$ or zero, depending on whether $T$ takes the value $t$ at a given 
trajectory or not. On the other hand, when taking the approach, that the notion of time (clocks)
should be provided by the dynamical fields, one finds, that the classical trajectory admits a unique
analytic extension, which completes it to entire $T\in\re$ \cite{ap-posL}.

The apparent difference between the results obtained in the Schr\"odinger picture and the group 
averaging approach has the following reason. In the former approach the choice of the particular 
self-adjoint extension corresponds to supplying an additional data into the system: the reflective
condition in $v=\infty$ (see \cite{ap-posL}). This allows to deterministically extend the evolution to 
all $T$. On the other hand the latter approach, by its very definition, we avoid supplying this additional 
data, instead evolving ``all the possibilities'' at once. That leads to the loss of 
of completeness of the set of observables $f(\hat{v})^{\rm D}_{(T,t)}$ given by functions $f$ of the  operator $\hat{v}$. That means, that from a set of expectation values
\be 
  \langle f(\hat{v})^{\rm D}_{T,t} \rangle 
\ee
at a given $t$, we can not predict the values 
\be 
  \langle f(\hat{v})^{\rm D}_{T,t'} \rangle \ , \qquad  t'\not= t \ . 
\ee
However one should stress, that the evolution of each observable $\hat{G}^{\rm D}_{(T,t)}$ 
in $t$ is unitary. The issue of the completeness loss will be addressed in detail in the future work.

\section{Discussion}\label{sec:concl}

In this paper we considered a quantum theory with a general constraint operator of the form
\begin{equation}
  {\hat C}\ =\ -\frac{\partial^2}{\partial T^2} \otimes
  \hat{B}-\hat{\id}\otimes \hat{H}.
\end{equation}
encountered for example in LQC. The issue we addressed was the uniqueness
and the properties of physical Hilbert spaces and observables one can define
within such theories. For that purpose we compared two constructions:
\begin{enumerate}[(i)]
  \item the Schr\"odinger  evolution picture used for example
    in \cite{APS-imp}, in which the constraint is
    reinterpreted as  (\ref{3a})  and
  \item the systematic procedure using the spectral decomposition of the contraint operator, a special case of the group averaging.
\end{enumerate}
Due to different mathematical properties of the (parametrized by
$\Pi\in\R$) operators $\hat{C}_{\ppi}:=\frac{1}{2}\ppi^2\hat{B}+\hat{C}_{\gr}$ we
restricted the comparison to two cases in which  spectra of $\hat{C}_{\ppi}$ are, respectively, discrete and
absolutely continuous. Moreover, in the continuous case we  assumed the asymptotic properties of the eigenfunctions which hold for example in the
LQC model the FRW spacetime coupled with the massless scalar field at the presence
of positive cosmological constant.

In the discrete case the physical Hilbert spaces and the evolution
picture (provided by constructed family of partial observables) for both listed
procedures coincide. In consequence for that case specified methods are
equivalent.

The situation changes in the continuous case. There, according to the Schr\"odinger 
picture the evolution is not unique, as the evolution operator $\hat{B}^{-1}\hat{C}_{\gr}$ 
(the square root of which plays there the role of a true Hamiltonian) admits inequivalent 
self-adjoint extensions. On the other hand the constraint operator $\hat{C}$ still remains essentially 
self-adjoint, thus the group averaging provides us with a unique (up to standard 
ambiguities tied to the procedure) Hilbert space. The comparison of this space with 
the ones corresponding to particular self-adjoint extensions in the Schr\"odinger picture 
shows that it is in a certain integral sense \eqref{eq:cont-descr} a union of all of the 
extensions. In consequence the physical evolution resulting from the group averaging can 
be understood as evolving all the extensions present in the Schr\"odinger picture at once 
(in parallel). Since the evolution picture defined in Section \ref{Exp2evolution} does not 
mix the subspaces corresponding to particular extensions, at least at this level they seem 
to look as the superselection sectors.

The situation complicates however, when we consider the observables. In the discrete 
case the construction following from GA and specified in Section \ref{Exp1} leads to operators, 
which coincide with the analogous operators constructed for the Schr\"odinger picture via method 
following from the initial value formulation (see \cite{APS-imp}). Both pictures would then predict 
exactly the same dynamics of a given physical system. In the continuous case however both the pictures start 
to differ. In the Schr\"odinger one all the observables are constructed on each of the Hilbert spaces 
corresponding to particular extensions of the evolution operator separately. Therefore they do not
mix these extensions by definition. On the other hand the GA construction results here in the 
operators, whose action mixes the subspaces corresponding to the extensions, while the operators 
corresponding to the different times are still related via unitary transformations.
Therefore pointed subspaces cannot be considered the superselection sectors anymore. Furthermore the presence 
of the 'nondiagonal' -- 
extension mixing terms in the operators might in principle indicate a possible difference in the 
dynamical predictions. This problem however has to be addressed in the context of particular examples, as 
the answer may strongly depend on the particular form and detailed spectral properties of the involved 
operators.

\section*{Acknowledgments} We have profited from discussions with Abhay
Ashtekar, Benjamin Bahr, Jan Derezi\'nski, Bianca Dittrich, Don Marolf, and Thomas Thiemann. This
work was supported in part by the NSF grant PHY0854743, the Polish Ministerstwo
Nauki i Szkolnictwa Wyzszego grants 1 P03B 075 29, 182/N-QGG/2008/0, the
2007-2010 research project N202 n081 32/1844 and Spanish MICINN Project
FIS2008-06078-C03-03, the Foundation for Polish Science grant Master, The
George A. and Margaret M. Downsbrough Endowment and the Eberly research funds of
Penn State. T.P. acknowledges also the financial aid by the I3P Program of
CSIC and the European Social Fund, and the funds of the European Research Council 
under short visit grant 3024 of GR network.

\appendix
\section{Continuous spectral decomposition} \label{ma_ap}

Here we briefly sketch the methods of singling out the space ${\cal H}_{{\rm
kin}\rho_0}$, which are presented in detail in \cite{Maurin}. For that,
let us introduce an commutative Von Neumann algebra $\cal W$ that is an intersection of commutant and double commutant of the constraints. Now we need to choose an $C^*$ algebra ${\cal A}\subset {\cal W}$, a dense subspace ${\mathcal D}\subset {\cal H}$ and a state $\mu$ on $\cal A$. Since our algebra is commutative one can identify it with an algebra of continuous functions on some compact space $K$ and state with a measure on it. We assume that the following conditions are satisfied:
\begin{itemize}
 \item $\cal A$ is separable and its weak closure is equal to $\cal W$.
\item For a pair $\phi,\phi'\in {\mathcal D}$ let us define a complex measure $\mu_{\phi,\phi'}$ on $K$ by
\be
  \mu_{\phi,\phi'}(A)=\langle \phi,\, A\, \phi'\rangle \ ,\quad A\in {\cal A} \ .
\ee
We assume that $\mu_{\phi,\phi'}$ is absolutely continuous with repsect to $\mu$ and its Radon-Nikodym derivative is a continuous function on $K$.
\end{itemize}
This unambiguously define Hilbert spaces in direct integral. However construction depends on the choice made. Measure theoretic version in choice independent.

In case of commuting quantum constraints $\hat{C}_I$, $I=1,...,d$ we take as $\cal A$ an algebra of bounded continuous functions of $\hat{C}_I$. In cases considered in this paper in some neighbourhood of $\rho_0$ all ${\cal H}_{{\rm
kin}\rho}$ will be isomorphic. In such a case, it accounts for the choices we made.

In the case we are considering 
\be 
  {\mathcal D}\ =\ {\rm span} \{\delta_v\ :\ v\in \R\}\otimes \ C^{\infty}_0(\R) \ , 
\ee
and it also have a natural structure of a nuclear space \cite{Maurin}.
We introduce also a notion
\be 
  {\cal D}_{gr}\ =\ {\rm span} \{\delta_v\ :\ v\in \R\} \ ,
\ee
that is also a nuclear space (as an limit of finite dimensional Hilbert spaces).

We assume some continuity of the Dirac observables $\hat{F}^{\rm D}$. Namely, we assume the continuity of $\hat{F}^{\rm D}(\psi_\rho)_\rho$ with respect to $\rho$ on the vectors from ${\mathcal D}$ i.e
distribution kernel of $F$ should be a continuous function. By formal definition we can take
absolute continuity of the measure
\be
  {\cal A}\otimes{\cal A}\ni A\otimes\, A'\ \rightarrow\  \left\langle \phi,\, A\hat{F}^{\rm
  D}A'\, \phi'\right\rangle
\ee
defined on $K\times K$ with respect to $\mu\times\mu$. We assume Radon-Nikodym derivative to be continuous.


\begin{thebibliography}{99}
  \bibitem{Marolf2} D.~Marolf, Refined algebraic quantization: Systems with a single
    constraint, (1995), \texttt{arXiv:gr-qc/9508015};\\
    D.~Marolf, {Quantum observables and recollapsing dynamics},
    Class.Quant.Grav. {\bf 12}, 1199-1220 (1995),
    \texttt{arXiv:gr-qc/9404053}.
  \bibitem{Marolf} 
    D.~Marolf, {Observables and a Hilbert space for Bianchi IX},
    Class.Quant.Grav. {\bf 12}, 1441-1454 (1995),
    \texttt{arXiv:gr-qc/9409049};\\
    D.~Marolf, {Almost ideal clocks in quantum cosmology: A brief
      derivation of time}, Class.Quant.Grav. {\bf 12}, 2469-2486 (1995),
    \texttt{arXiv:gr-qc/9412016};\\
    A.~Ashtekar, J.~Lewandowski, D.~Marolf,
    J.~Mour\~ao and T.~Thiemann, Quantization of diffeomorphism
    invariant theories of connections with local degrees of freedom,
    {J.Math.Phys.} \textbf{36}, 6456-6493 (1995),
    \texttt{arXiv:gr-qc/9504018}.
  \bibitem{Maurin} K.~Maurin, {\em General eigenfunction expansions and unitary 
    representations of topological groups} (PWN, Warsaw, 1968).
  \bibitem{Rovelli1}
    C.~Rovelli, {Partial observables},
    Phys.Rev. {\bf D65} 124013 (2002), \texttt{arXiv:gr-qc/0110035}.
  \bibitem{rel-obs} C.~Rovelli, {WHAT IS OBSERVABLE IN CLASSICAL AND QUANTUM GRAVITY?},
    Class.Quant.Grav. {\bf 8} 297 (1991);\\
    B.~Dittrich, {Partial and complete observables for Hamiltonian
      constrained systems},
    Gen.Rel.Grav. {\bf 39} 1891 (2007), \texttt{arXiv:gr-qc/0411013};\\
    B.~Dittrich, {Partial and Complete Observables for Canonical
      General Relativity},
    Class.Quant.Grav. {\bf 23} 6155 (2006), \texttt{arXiv:gr-qc/0507106};\\
    T.~Thiemann, {Reduced phase space quantization and Dirac observables},
    Class.Quant.Grav. {\bf 23} 1163 (2006), \texttt{arXiv:gr-qc/0411031}.
  \bibitem{Rovelli} C.~Rovelli, {\em Quantum Gravity}
    (CUP, Cambridge, 2004).
  \bibitem{A-LQG}
    A.~Ashtekar and J.~Lewandowski, {Background Independent Quantum
    Gravity: A Status Report}, Class.Quant.Grav. \textbf{21}, R53 
    (2004), \texttt{arXiv:gr-qc/0404018}.
  \bibitem{Thiemann} T.~Thiemann, {\em Introduction to Modern Canonical
    Quantum General Relativity} (CUP, Cambridge, 2007).
  \bibitem{lqg-bk} K.~Giesel and T.~Thiemann, {\em Algebraic Quantum Gravity 
    (AQG) IV. Reduced Phase Space Quantisation of Loop Quantum Gravity},    
    (2007), \texttt{arXiv:0711.0119}.
  \bibitem{LQC}
    M.~Bojowald, {Loop Quantum Cosmology}, Living
    Rev.Rel. \textbf{8}, 11 (2005), \texttt{arXiv:gr-qc/0601085}.
  \bibitem{A-LQC}
    A.~Ashtekar, {An Introduction to Loop Quantum Gravity Through
    Cosmology}, Nuovo Cim. \textbf{122B}, 135-155 (2007),
    \texttt{arXiv:gr-qc/0702030}.
  \bibitem{LQC-ABL}
    A.~Ashtekar, M.~Bojowald, J.~Lewandowski,
    {Mathematical structure of loop quantum cosmology}, Adv.Theo.%
    Math.Phys. \textbf{7}, 233-268 (2003), \texttt{arXiv:gr-qc/0304074}.  
  \bibitem{klp-aspects} W.~Kami\'nski, J.~Lewandowski and T.~Paw{\l}owski, 
    Physical time and other conceptual issues of QG on the example of LQC,
    Class.Quant.Grav. {\bf 26}, 035012 (2009) \texttt{arXiv:0809.2590}.
  \bibitem{kp-constr}
    L.~Szulc, W.~Kami\'nski and J.~Lewandowski, 
    {Closed FRW model in Loop Quantum Cosmology}, Class.Quant.Grav. 
    \textbf{24}, 2621-2635 (2007), \texttt{arXiv:gr-qc/0612101};\\
    W.~Kami\'nski and J.~Lewandowski, 
    {The flat FRW model in LQC: the self-adjointness}, (2007),
    \texttt{arXiv:0709.3120}.
  \bibitem{APS} A.~Ashtekar, T.~Paw{\l}owski and P.~Singh, {Quantum nature
      of the big bang: An analytical and numerical investigation},
    Phys.Rev. {\bf D73}, 124038 (2006), \texttt{arXiv:gr-qc/0604013}.
  \bibitem{APS-imp} A.~Ashtekar, T.~Paw{\l}owski, and P.~Singh, 
    {Quantum nature of the big bang: Improved dynamics},
    Phys.Rev. {\bf D74}, 084003 (2006), {\texttt{arXiv:gr-qc/0607039}.}
  \bibitem{ap-posL} A.~Ashtekar , T.~Paw{\l}owski, 
    Loop quantum cosmology and the positive cosmological constant,
    {\it in prep.}
  \bibitem{kp-posL} W.~Kami\'nski , T.~Paw{\l}owski, The LQC evolution operator of FRW
    universe with positive cosmological constant,
    {\it in prep.}
  \bibitem{bp-negL} E.~Bentivegna and T.~Paw{\l}owski, {Anti-deSitter universe dynamics in LQC},
        Phys.Rev. {\bf D77}, 124025 (2008), {\texttt{arXiv:0803.4446}.}
  \bibitem{apsv} A.~Ashtekar, T.~Paw{\l}owski, P.~Singh and
    K.~Vandersloot, {Loop quantum cosmology of k=1 FRW models},
    Phys.Rev. {\bf D75}, 024035 (2007), \texttt{arXiv:gr-qc/0612104}.
  \bibitem{b1-szulc}
    L-.~Szulc, {Loop Quantum Cosmology of Diagonal Bianchi Type I model: 
    simplifications and scaling problems}, Phys.Rev. {\bf D78}, 064035 (2008),
    \texttt{arXiv:0803.3559}.
  \bibitem{Fulling} S.A.~Fulling, {\em Aspects of quantum field theory in curved space-time}
    (CUP, Cambridge, 1989).
  \bibitem{mmp2}
    M.~Mart\'in-Benito, G.A.~Mena-Marug\'an and T.~Paw{\l}owski, 
    {Physical evolution in Loop Quantum Cosmology: The example of vacuum Bianchi I}, (2009),
    \texttt{arXiv:0906.3751}.
\end{thebibliography}
\end{document}